\documentclass{aa}  

\usepackage{graphicx}
\usepackage{caption}
\usepackage{subcaption}
\usepackage{hyperref}
\usepackage{txfonts}
\usepackage{xcolor}
\usepackage{soul}  
\setstcolor{red}

\usepackage{lineno}
\let\linenumbers\relax
\let\nolinenumbers\relax

\allowdisplaybreaks[1]

\DeclareMathOperator{\Tr}{Tr}

\bibpunct{(}{)}{;}{a}{}{,}

\begin{document} 
\title {Fast magneto-acoustic waves in the solar chromosphere: Comparison of single-fluid and two-fluid approximations.}

\author{M.M. Gómez-Míguez \inst{1,2} \and
           D. Martínez-Gómez \inst{1,2} \and
           E. Khomenko \inst{1,2} \and
           B. Popescu Braileanu \inst{3} \and
           M. Collados \inst{1,2} \and
           P.S. Cally \inst{4}
          }

\institute{Instituto de Astrofísica de Canarias, 38205 La Laguna, Tenerife, Spain
              \and
             Departamento de Astrofísica, Universidad de La Laguna, 38205, La Laguna, Tenerife, Spain
             \and
             Department of Physics and Technology, University of Bergen, Bergen, Norway
             \and
             School of Mathematics, Monash University Victoria, 3800, Australia
             }

 
  \abstract
   {
   The mechanism behind the heating of the solar chromosphere remains unclear. Friction between neutrals and charges is expected to contribute to plasma heating in a partially ionised plasma (PIP).}
   {We aim to study the efficiency of the frictional heating mechanism in partially ionised plasmas by comparing a single-fluid model (1F) using ambipolar diffusion and a two-fluid model (2F) that incorporates elastic collision terms.}
   {We use the code MANCHA-2F to solve the equations for both models numerically. The simulations involve the vertical propagation of fast magneto-acoustic waves from the top of the photosphere through the chromosphere. The model atmosphere is vertically stratified, including a horizontal, homogeneous magnetic field. We also apply the linear theory to supplement the numerical results. Finally, we look at the assumptions of the 1F model to find out what causes the model discrepancies.}
   {The results show that the temperature increase for the 1F model is slightly higher than for the 2F model, especially with long-period waves. The wave energy flux indicates that in the 2F model, the wave is transporting less energy upwards. From the linear theory, we find that the wave in the 2F model loses more energy than in the 1F model in the deep layers, but the opposite occurs in the high layers.}
   {The efficient dissipation in the 2F model in deep layers reduces the energy flux at the high layers, reducing heating and explaining the temperature differences between models. We attribute those discrepancies to the contribution of pressure forces to the drift velocity and the omission of a term related to the centre of mass frame reference in the 1F energy equation.} 
   

   \maketitle
%
\section{Introduction \label{sec:I}}

Owing to the relatively cool temperatures in the solar photosphere and chromosphere, the plasma in these layers (primarily consisting of hydrogen) is partially ionised.
As one moves to upper layers, the magnetic pressure exceeds the gas pressure and the magnetic field determines the behaviour of charges. 
Neutral particles only sense the presence of the magnetic field through collisions with charges. The friction resulting from charge-neutral collisions causes the dissipation of kinetic and magnetic energies, which then convert into internal energy, raising the plasma temperature. This process becomes more effective when both the magnetohydrodynamic (MHD) and collision timescales are comparable \citep{ZaqKhoRuc2011aa, SolCarBal2013aa, SolCarBal2013ab, BalAleCol2018aa, PopKep2021aa}. \citet{KhoArbRuc2004aa,KhoRucOli2006aa} demonstrated that, under conditions observed in prominences, wave damping resulting from charge-neutral collisions is more pronounced than damping caused by viscosity or thermal conductivity. Furthermore, \citet{SheKhoVic2016aa} found that charge-neutral collisional damping is more efficient than Ohmic dissipation for propagating waves in flux tubes.

Depending on the scales, describing the interaction between charges and neutrals might require a model that includes both components as separated fluids. A reference work for the description of a multi-fluid plasma (N-fluid, or NF) is the seminal paper of Braginskii \citep{Bra1965aa}, among other works \citep[see, e.g.][]{ChaCow1970aa, SchNag2009aa}. This approach entails computing the moments of the Boltzmann equation for the distribution function of each component of the plasma. The Chapman-Enskog successive-approximation technique calculates the high-order moments of the distribution \citep{Sch1977aa,MeiShu2012aa}. Braginskii's method applies to highly collisional plasmas and is valid across most of the chromosphere. For plasmas with intermediate or low collision coupling, a more general model, such as the one developed by \citet{HunTenZan2019aa} and \citet{HunPasKho2022aa}, is required. Recently, \citet{Hun2024aa} extended Braginskii's set of equations to a 22-moment model for a 2F plasma composed by neutrals and charges.

When dealing with partially ionised solar plasmas, a frequently used method for simulating the interaction between charged and neutral particles is using a single-fluid model (1F) \citep[for example][]{ZweBra1997aa, BalAleCol2018aa, NobMarMor2020aa}. This model includes ambipolar diffusion among other non-ideal terms in the generalised Ohm's law. Outside of solar physics, it has been used, for example, in simulating neutron stars \citep{Jon1987aa}, the interstellar medium \citep{Bra2019aa} or weakly ionised protoplanetary disks \citep{BaiSto2011aa}. Many numerical studies in solar physics have invoked the ambipolar term. For instance, \citet{CheCam2012aa}, \citet{MarDePHan2012aa} and \citet{KhoVitCol2018aa} found ambipolar diffusion to produce measurable effects in magneto-convection simulations, exceeding artificial diffusivities at chromospheric layers. \citet{MarDePCar2016aa} analysed the misalignment of magnetic field lines with the visible structure of spicules and due to the interaction between charges and neutrals, invalidating the MHD assumption that plasma motion keeps following the same magnetic
field lines. \citet{NobMorMar2020aa} and
\citet{MarLeeDeP2020aa,MarCruGos2023aa} obtained that nonequilibrium (NEQ) conditions may affect the efficiency of ambipolar diffusion. Recently, \citet{GoMMarKho2024aa} incorporated inelastic collisions into ambipolar diffusion term definition, finding the corrections to the ambipolar coefficient to be negligible in quiet Sun conditions. The works by
\citet{KhoArbRuc2004aa},
\citet{KhoRucOli2006aa},
\citet{ForOliBal2007aa}, \citet{SonVas2011aa}, 
\citet{SolCarBal2015aa}, \citet{SheKhoVic2016aa}, \citet{CalKho2019aa}, \citet{KhoCal2019aa}, or
\citet{PopLukKho2021aa}, among others, have studied the damping of MHD waves and currents due to ambipolar diffusion. They found high-frequency Alfvén and fast waves to be effectively damped through this mechanism in many different scenarios such as quiet Sun, flux tubes, or prominences. However, for the case of slow waves, \citet{ForOliBal2007aa} found them to be weakly affected by ion-neutral collisions. Moreover, \citet{TerMolWie2002aa}, \citet{TerCarOli2005aa}, \citet{CarTerOli2006aa}, \citet{ForOliBal2008aa}, or  \citet{SolCarBal2015aa} found them to be more affected by other non-ideal mechanisms, such as viscosity or thermal conduction.

Over the past years, many NF studies have appeared in the context of Solar Physics. On wave theory, \citet{ZaqKhoRuc2011aa, ZaqKhoSol2013aa} and \citet{SolCarBal2013aa, SolCarBal2013ab} characterized the 2F normal modes in an uniform atmosphere, focusing on the cutoff regions not present in the 1F approach \citep[see, e.g.,][]{KulPea1969aa, KamNis1998aa}. Expanding the analysis of 2F waves, \citet{Cal2023aa} studied the 10 modes resulting from the linear equations applied to a homogeneous atmosphere, focusing on the efficiency in generating those modes and  \citet{CalGoM2023aa} have applied ray tracing formalism for the study of fast-to-slow mode conversion in 2F under the assumption of weak dissipation. We refer to the review by \citet{BalAleCol2018aa} and \citet{Sol2024aa} for recent developments in 2F waves.

Regarding numerical simulations, \citet{PopLukKho2019aa, PopLukKho2019ab} or \citet{ZhaPoeLan2021aa} have studied wave heating in a 1D stratified atmosphere for different wave-periods, finding the temperature increments to decrease for shorter wave-periods. \citet{SnoHil2021aa} delved into 1D shock propagation, finding neutrals to stabilise the wavefront for finite collisions.
Furthermore, \citet{Mei2011aa}, \citet{ManAlvLan2017aa}, \citet{Bal2019aa}, \citet{ZhaPoeLan2021aa} have expanded the research by incorporating ionisation-recombination effects, revealing challenges in configuring a static background atmosphere when hydrostatic equilibrium is imposed. Recently, \citet{SnoDruHil2023aa} also included radiative ionisation-recombination rates for a (5 + 1 levels) hydrogen atom for investigating shock propagation and \citet{PopKep2024aa} added radiative cooling under coronal conditions for studying the growth of the thermal instability and acoustic wave damping. Apart from the 2F modeling, \citet{MarSolTer2017aa} have explored a 5F approach that includes hydrogen and helium, and \citet{MarSolTer2018aa} have used a 3F model to study the non-linear regime of Alfvén waves in a 1D homogeneous atmosphere. In addition, the code Ebysus \citep{MarSzyHan2020aa} is capable of performing NF simulations for different species, treating separately ionisation and excitation states.

Although NF models offer greater flexibility, make fewer assumptions and are, therefore, a more accurate description of chromospheric plasma at small scales, it is still possible that a simpler 1F model is sufficiently precise for application in the chromosphere, depending on the frequency of the studied process. Hence, it is necessary to compare both approaches and verify the applicability of the 1F approach under conditions similar to those of the solar chromosphere. This study aims to enhance our understanding of the heating mechanisms associated with interactions between charged particles and neutrals and to determine the specific scenarios where the 1F approximation becomes less accurate than the more detailed 2F approximation. For this purpose, we employed the MANCHA-2F code developed by \citet{PopLukKho2019aa} to solve both the 1F and 2F equations and compare their solutions in a simple scenario involving the vertical propagation of fast magneto-acoustic waves in a solar-like chromospheric model, following the methodology of previous research \citep[e.g.,][]{PopLukKho2019ab}.

The paper is organised as follows. Section~\ref{sec:theo} outlines the mathematical models utilised in this study; Section~\ref{sec:simulations} presents the results of the numerical experiments on the propagation of fast magneto-acoustic waves; Section~\ref{sec:eikonal} presents the analytical solution for the wave equation; Section \ref{sec:discrep} analyses the reasons for the 1F and 2F discrepancies and, finally, Section~\ref{sec:Conclusions} summarises our findings.


\section{Physical description of the plasma \label{sec:theo}}

In this work, we assume a hydrogen plasma consisting of electrons, protons, and neutral particles, without other species, molecules or negative hydrogen ions. We focus only on the effects due to elastic charge-neutral collisions. We do not include inelastic collisions such as ionisation or recombination and do not consider viscosity or heat flux. By restricting ourselves to a 5-moment model \citep{Sch1977aa}, we can consistently apply these assumptions in a closed model. Ohmic heating, Hall effect or Biermann battery effect are not included in the Generalized Ohm's law.

\subsection{The two-fluid (2F) model\label{sec:2F}}

Our model follows a 2F approach, where elastic collisions couple a neutral fluid with a charge fluid. The neutral fluid has only one component, while the charge fluid consists of strongly coupled protons and electrons. In this instance, we are examining a 2F model comprising charges and neutrals, as established by \citet{MeiShu2012aa}, \citet{LeaLukLin2013aa}, or \citet{Pop2020aa}. The set of moment equations is

\begin{subequations}
\begin{gather}
\frac{\partial \rho_{\rm n}}{\partial t} + \nabla \cdot (\rho_{\rm n} \vec{u}_{\rm n}) = 0, \label{eq:rho_n} \\
\frac{\partial \rho_{\rm c}}{\partial t} + \nabla \cdot (\rho_{\rm c} \vec{u}_{\rm c}) = 0, \label{eq:rho_c}\\
\frac{\partial}{\partial t}(\rho_{\rm n} \vec{u}_{\rm n}) + \nabla \cdot (\rho_{\rm n} \vec{u}_{\rm n} \otimes \vec{u}_{\rm n} + \hat{p}_{\rm n})~~~~~~~~ = \rho_{\rm n} \vec{g} + \vec{R}_{\rm n}, \label{eq:mom_n}\\ 
\frac{\partial}{\partial t}(\rho_{\rm c} \vec{u}_{\rm c}) + \nabla \cdot (\rho_{\rm c} \vec{u}_{\rm c} \otimes \vec{u}_{\rm c} + \hat{p}_{\rm c} + \hat{p}_{\rm m}) = \rho_{\rm c} \vec{g} - \vec{R}_{\rm n}, \label{eq:mom_c}\\ 
\frac{\partial e_{\rm n}}{\partial t} + \nabla \cdot \left[e_{\rm n}\vec{u}_{\rm n} + \hat{p}_{\rm n} \cdot \vec{u}_{\rm n} \right] ~~~~~~~~~~~ =  \rho_{\rm n} \vec{g} \cdot \vec{u}_{\rm n} + M_{\rm n}, \label{eq:eq_e_n} \\
\frac{\partial e_{\rm c}}{\partial t} + \nabla \cdot \left[e_{\rm c} \vec{u}_{\rm c} + \left(\hat{p}_{\rm c} + \hat{p}_{\rm m}\right)\cdot \vec{u}_{\rm c} \right] = \rho_{\rm c} \vec{g} \cdot \vec{u}_{\rm c} - M_{\rm n}, \label{eq:eq_e_c}
\end{gather}
\end{subequations}

\noindent
where the usual thermodynamic variables $\rho_\alpha,~\vec{u}_\alpha,~\hat{p}_\alpha,~e_{\rm \alpha}$ are, respectively, the mass density, the centre of mass velocity, the stress tensor and the total energy of the fluid $\alpha ={\rm c, n}$ for charges and neutrals, respectively. Note that the charges include the sum of protons and electrons. The scalar product is denoted by `$\cdot$' and the outer product by `$\otimes$'. The total energy is, in each case,
\begin{equation}
e_{\rm n} = e_{\rm int,n} + \frac{1}{2} \rho_{\rm n} u_{\rm n}^2;~~~ e_{\rm c} = e_{\rm int,c} + \frac{1}{2} \rho_{\rm c} u_{\rm c}^2 + p_{\rm m};~~~ e_{\rm int,\alpha} = \frac{p_\alpha}{\gamma - 1} \label{eq:totealpha}
\end{equation}
with $e_{\rm int,\alpha}$ the internal energy and the adiabatic index $\gamma = 5 / 3$. Density and pressure obey the ideal gas relations 
\begin{subequations}
\begin{gather}
    p_{\rm n} = \frac{\rho_{\rm n}}{m_H}k_{\rm B} T_{\rm n}, \label{eq:p_n}\\
    p_{\rm c} = \frac{2\rho_{\rm c}}{m_{\rm H}}k_{\rm B} T_{\rm c}, \label{eq:p_c}
\end{gather}
\end{subequations}
where $p_\alpha = \frac{1}{3}\Tr{\hat{p}_\alpha}$ is the scalar pressure, $m_{\rm H}$ the mass of hydrogen atom, $k_{\rm B}$ the Boltzmann constant and $T_\alpha$ the temperature. 

The gravitational acceleration \vec{g} and the magnetic field $\vec{B}$ characterize the macroscopic forces acting on the system. The magnetic force is described throughout the divergence of the magnetic stress tensor, $\hat{p}_m$, defined as:
\begin{subequations}
\begin{gather}
    \hat{p}_m = p_m \hat{I} - \frac{1}{\mu_0}\vec{B} \otimes \vec{B},\label{eq:tens_p_m}\\
    p_m = \frac{B^2}{2\mu_0}, 
    \label{eq:p_m} 
\end{gather}
\end{subequations}
where $\mu_0$ is the vacuum magnetic permeability, $p_{\rm m}$ is the magnetic pressure and $\hat{I}$ is the 3 $\times$ 3 identity tensor. The magnetic force is commonly expressed in terms of the current density $\vec{J} = \frac{1}{\mu_0} \vec{\nabla} \times \vec{B}$, so $\nabla \cdot \hat{p}_{\rm m} = - \vec{J} \times \vec{B}$.

The magnetic field evolves according to the induction equation
\begin{equation}
    \frac{\partial \vec{B}}{\partial t} + \nabla \cdot (\vec{u}_{\rm c} \otimes \vec{B} - \vec{B} \otimes \vec{u}_{\rm c}) = 0, 
    \label{eq:induction} 
\end{equation}
and obeys the no monopole condition
\begin{equation}
    \nabla \cdot \vec{B} = 0. \label{eq:no_monopole}
\end{equation}

The terms $\vec{R}_{\rm n}$ and $M_{\rm n}$ describe the elastic collisional interaction between the fluids \citep{Bra1965aa}. The elastic collision term for the mass equation is null since no particles are produced or destroyed in elastic processes. For the case of a hydrogen plasma, the collisional terms are given by
\begin{gather}
    \vec{R}_{\rm n} = \alpha \rho_{\rm c} \rho_{\rm n} (\vec{u}_{\rm c} - \vec{u}_{\rm n}), \label{eq:R_n}\\
    M_{\rm n} = \frac{1}{2}\alpha \rho_{\rm c} \rho_{\rm n} (u_{\rm c}^2 - u_{\rm n}^2) + \frac{1}{\gamma  - 1}\frac{k_{\rm B}}{m_{\rm H}}\alpha \rho_{\rm c} \rho_{\rm n} (T_{\rm c} - T_{\rm n}), \label{eq:M_n}
\end{gather}
where $\alpha$ is the collision parameter:
\begin{equation}
    \alpha(T_{\rm eff}) = \frac{1}{m_{\rm H}^2}\sqrt{\frac{8 k_{\rm B} T_{\rm eff}}{\pi}}(\sqrt{m_{\rm pn}}\, \sigma_{\rm pn} + \sqrt{m_{\rm en}}\, \sigma_{\rm en}).
    \label{eq:coll_param}
\end{equation}
The subindex ``p'' is for protons, 
$m_{\rm \beta n} = m_\beta m_{\rm n} / (m_\beta + m_{\rm n})$, $T_{\rm eff} = (T_{\rm c} + T_{\rm n}) / 2$ and $\sigma_{\rm \beta n} = 10^{-19} \ \rm{m^{2}}$ are the cross-sections, assuming hard sphere elastic collisions \citep{DraRobDal1993aa,LeaLukLin2012aa}, for $\beta = \rm{p,e}$. By using these definitions, the collision frequencies are,
\begin{subequations}
\begin{gather}
    \nu_{cn} = \alpha \rho_n, \label{eq:nu_cn}\\
    \nu_{nc} = \alpha \rho_c.
    \label{eq:nu_nc}
\end{gather}
\end{subequations}

These frequencies define the collision time scales $\tau_{\rm cn} = 1 / \nu_{\rm cn}$ and $\tau_{\rm nc} = 1 / \nu_{\rm nc}$. If the MHD time scale of the studied processes is similar to or smaller than the collision ones, the 2F nature of the plasma arises.

\subsection{A single fluid (1F) model with ambipolar diffusion \label{sec:1F}}
Following the same notation as in the 2F section \ref{sec:2F}, the 1F set of equations reads as 
\begin{subequations}
\begin{gather}
    \frac{\partial \rho}{\partial t} + \nabla \cdot (\rho \vec{u}) = 0, \label{eq:rho}\\
    \frac{\partial}{\partial t}(\rho \vec{u}) + \nabla \cdot (\rho \vec{u} \otimes \vec{u} + \hat{p} + \hat{p}_{\rm m}) = \rho \vec{g}, \label{eq:mom}\\
    \frac{\partial e}{\partial t} + \nabla \cdot \left[e \vec{u} + (\hat{p} + \hat{p}_{\rm m}) \cdot \vec{u} + \vec{S}^* \right] = \rho \vec{g} \cdot \vec{u},  \label{eq:eq_e} \\
    \frac{\partial \vec{B}}{\partial t} + \nabla \cdot (\vec{u} \otimes \vec{B} - \vec{B} \otimes \vec{u}) = - \nabla \times \vec{E}^*, \label{eq:induc_1F}
\end{gather}
\end{subequations}
with the thermodynamical realtions,
\begin{subequations}
  \begin{gather}
    e = e_{\rm int} + \frac{1}{2}\rho u^2 + p_{\rm m};~~~ e_{\rm int} = \frac{p}{\gamma - 1} \label{eq:e}\\    
    p = \frac{\rho}{\tilde{\mu} m_{\rm n}} k_{\rm B} T, \label{eq:p}\\
    \tilde{\mu} = \frac{1}{2 - \xi_{\rm n}}, \label{eq:mu}\\
    \xi_{\rm c} + \xi_{\rm n} = \frac{\rho_{\rm c}}{\rho} + \frac{\rho_{\rm n}}{\rho} = 1. \label{eq:xi}
  \end{gather}  
\end{subequations}

The amount of neutrals and charges is encapsulated on the mass fraction, $\xi_\alpha$, and on the average particle mass, $\tilde{\mu}$.

In the 1F approach, ambipolar diffusion is included through the Generalized Ohm's law. This effect is present in Eqs. \eqref{eq:eq_e} and \eqref{eq:induc_1F}
by introducing the non-ideal electric field $\vec{E}^*$ and the Poynting flux $\vec{S}^*$, which can be expressed as
\begin{subequations}
\begin{gather}
    \vec{E}^* = - \eta_{\rm A} \frac{(\vec{J} \times \vec{B}) \times \vec{B}}{B^2} = - \tilde{\eta}_{\rm A} \vec{B} \times \nabla \cdot \hat{p}_{\rm m}, \label{eq:Estar}\\
    \vec{S}^* = \frac{1}{\mu_0} \vec{E}^* \times \vec{B} = -
    \frac{\eta_{\rm A}}{\mu_0}  \frac{(\vec{J} \times \vec{B}) \times \vec{B}}{B^2} \times \vec{B} =\label{eq:Sstar}\\~~~~~~~~
    - \frac{1}{\mu_0} (\tilde{\eta}_{\rm A} \vec{B} \times \nabla \cdot \hat{p}_{\rm m}) \times \vec{B}, \nonumber
\end{gather}
\end{subequations}
where we express the ambipolar coefficient as,
\begin{equation}
    \tilde{\eta}_{\rm A} = \frac{\xi_{\rm n}^2}{\alpha \rho_{\rm c} \rho_{\rm n}}.
\label{eq:eta_A}
\end{equation}

We also introduce the more usual ambipolar coefficient $\eta_{\rm A} = B^2 \tilde{\eta}_{\rm A}$ and express the ambipolar terms in a more standard fashion. However, we will make use of the non-standard notation in the following sections.

The momentum exchange terms in the 2F (see Eq. \eqref{eq:R_n}) and the ambipolar diffusion refer to the same process: the diffusion of neutrals and charges. It is worthwhile to clarify the connection between these two concepts. Here, we follow the steps outlined by, namely, \citet{KhoColDia2014aa}.

We consider a 1F model that accounts for velocity drifts between neutrals and charges,
\begin{equation}
    \vec{w} = \vec{u}_{\rm c} - \vec{u}_{\rm n}. \label{eq:w}
\end{equation}

In this approach, the dynamics of the plasma is described by the centre of mass of the components, which moves at the velocity
\begin{equation}
    \vec{u} = \frac{\rho_{\rm n} \vec{u}_{\rm n} + \rho_{\rm c} \vec{u}_{\rm c}}{\rho_{\rm n} + \rho_{\rm c}}.\label{eq:u_CM}
\end{equation}

With the above
considerations, we can relate the drift velocity of neutrals and charges respect to the centre of mass, $\vec{w}_{\alpha}$, with the drift velocity $\vec{w}$, by using Eq, \eqref{eq:w} and \eqref{eq:u_CM}, 
\begin{subequations}
    \begin{gather}
        \vec{w}_{\rm n} =\vec{u}_{\rm n} - \vec{u} = - \xi_{\rm c} \vec{w}, \label{eq:wn}\\
        \vec{w}_{\rm c} = \vec{u}_{\rm c} - \vec{u} = \xi_{\rm n} \vec{w}.
        \label{eq:wc}
    \end{gather}
\end{subequations} 

From the inspection of the charge energy equation and the 2F induction equation (see Eqs. \eqref{eq:eq_e_c} and \eqref{eq:induction}), it follows that the magnetic field terms are related to the velocity of the charged fluid. The non-ideal terms $\vec{E}^*$ and $\vec{S}^*$ (see Eqs. \eqref{eq:Estar} and \eqref{eq:Sstar}) result from the change of the charge velocity to the velocity of the centre of mass frame velocity, so
\begin{equation}
    \vec{u}_{\rm c} = \vec{u} + \xi_{\rm n}\vec{w}, \label{eq:uc_u}
\end{equation}
and,
\begin{equation}
        \vec{E}^* = \vec{E} + \vec{u} \times \vec{B} =  - \xi_{\rm n} \vec{w} \times \vec{B}. \label{eq:Estarw}
\end{equation}

The drift velocity $\vec{w}$ is not a native variable of the 1F model, as it requires to evolve the velocities for neutrals and charges separately. Therefore, it is necessary to make some assumptions in order to account for the drift velocity. By combining the momentum equations for neutrals and charges (Eqs. \eqref{eq:mom_n} and \eqref{eq:mom_c}), we find,
\begin{gather}
    \xi_{\rm n}\left(\frac{\partial}{\partial t}(\rho_{\rm c} \vec{u}_{\rm c}) + \nabla \cdot (\rho_{\rm c} \vec{u}_{\rm c} \otimes \vec{u}_{\rm c})\right) \nonumber\\
     -\xi_{\rm c}\left(\frac{\partial}{\partial t}(\rho_{\rm n} \vec{u}_{\rm n}) + \nabla \cdot (\rho_{\rm n} \vec{u}_{\rm n} \otimes \vec{u}_{\rm n})\right) =  - \xi_{\rm n} \nabla \cdot \hat{p}_{\rm m} - \vec{G} - \alpha \rho_{\rm c}\rho_{\rm n} \vec{w},\label{eq:fullw} 
\end{gather}
similar to Eq. (129) by \citet{KhoColDia2014aa}. The terms on the left hand side (LHS) are the inertial terms, which approximately read as,
\begin{gather}
    \xi_{\rm n}\left(\frac{\partial}{\partial t}(\rho_{\rm c} \vec{u}_{\rm c}) + \nabla \cdot (\rho_{\rm c} \vec{u}_{\rm c} \otimes \vec{u}_{\rm c})\right) \nonumber\\-\xi_{\rm c}\left(\frac{\partial}{\partial t}(\rho_{\rm n} \vec{u}_{\rm n}) + \nabla \cdot (\rho_{\rm n} \vec{u}_{\rm n} \otimes \vec{u}_{\rm n})\right)     \approx  \rho \xi_{\rm c} \xi_{\rm n} \frac{\partial \vec{w}}{\partial t}.\label{eq:inertia} 
\end{gather}

The ratio between the inertial terms and the collisional momentum exchange term  $\mathbf{R_n}$
can be roughly estimated by
\begin{equation}
    \varpi_{\rm inertia} = \frac{\omega}{\alpha \rho},
    \label{eq:Est_inertia}
\end{equation}
with $\omega$ the characteristic frequency of the process. 

If $\varpi_{\rm inertia} \ll 1$, the inertial terms can be neglected in Eq. \eqref{eq:fullw}, and then
\begin{equation}
    \vec{w}_{\rm 1F} = \frac{\tilde{\eta}_{\rm A}}{\xi_{\rm n}^2}(\xi_{\rm n} \vec{J} \times \vec{B} - \vec{G})= - \frac{\tilde{\eta}_{\rm A}}{\xi_{\rm n}^2}(\xi_{\rm n} \nabla \cdot \hat{p}_{\rm m} + \vec{G}).\label{eq:w_1F}
\end{equation}

The first term on the right-hand side (RHS) of Eq. \eqref{eq:w_1F} is the drift velocity due to magnetic forces, which is usually referred to as ambipolar diffusion \citep[see also the discussion by ][]{NobMarMor2020aa}. This contribution is included in the 1F model presented here through the terms $\vec{E}^*$ and $\vec{S}^*$ (see eqs. \eqref{eq:Estar} and \eqref{eq:Sstar}). The second one is the thermal pressure function, given by
\begin{equation}
    \vec{G} = \xi_{\rm n} \nabla \cdot \hat{p}_{\rm c} - \xi_{\rm c} \nabla \cdot \hat{p}_{\rm n}, \label{eq:G}
\end{equation}
which describes the drift due to the thermal pressure forces. In the absence of viscosity and assuming thermal equilibrium ($T_{\rm n} = T_{\rm c}$) and scalar pressures, it can be reduced to
\begin{equation}
    \vec{G} = \xi_{\rm n} (1 - \xi_{\rm n}) \,\nabla (\tilde{\mu} p) - (1 + \xi_{\rm n})\, \tilde{\mu}\, p \nabla \xi_{\rm n} ,\label{eq:G_novisc}
\end{equation}
which is equivalent to the expression obtained by \citet{ForOliBal2007aa}. In principle, we ignore this term in the 1F approach, since we study a weakly ionized atmosphere and $\vec{G} \to 0$ in both the limits of weak and strong ionisation \citep[see the discussion by ][]{KhoArbRuc2004aa, KhoRucOli2006aa, PinGalBac2008aa}. Additionally, in Eq \eqref{eq:w_1F}, there is a missing term proportional to the current density and to the electron mass \citep[see e.g.,][]{Bra1965aa, KhoColDia2014aa, BenFli2020aa}. This term comes from the drift velocity between ions and electrons and it is neglected in the 2F model discussed in this work \citep[for more details, see the model decription by][]{Pop2020aa}.

\subsection{Internal energy equation \label{sec:eint}}
The energy equation can be cast in several alternative forms, showing explicitly different physical properties of the system. The internal energy equation is specially useful because it is connected with the temperature throughout the state equation. In equations \eqref{eq:eq_e_n}, \eqref{eq:eq_e_c} and \eqref{eq:eq_e}, it is possible to remove the explicit contribution of the kinetic and magnetic energy by using the momentum and induction equations, obtaining for the 2F model:
\begin{subequations}
\begin{gather}
     \frac{\partial e_{\rm int,n}}{\partial t} + \nabla \cdot (e_{\rm int,n} \vec{u}_{\rm n}) = - p_{\rm n} \nabla \cdot \vec{u}_{\rm n} + Q_{\rm n} \label{eq:eq_e_n_in},\\
     \frac{\partial e_{\rm int,c}}{\partial t} + \nabla \cdot (e_{\rm int,c} \vec{u}_{\rm c}) = - p_{\rm c} \nabla \cdot \vec{u}_{\rm c} + Q_{\rm c}. \label{eq:eq_e_c_in}
\end{gather}
\end{subequations}
\noindent
Hereafter, we simplify the notation by using only the scalar pressure. The RHS terms represent the internal energy sources. The first one is the adiabatic source term and $Q_{\alpha}$ accounts for the energy variation due to elastic collisions. Explicitly, they are
\begin{subequations}
\begin{gather}
 Q_{\rm n} =\frac{1}{2}\alpha \rho_{\rm c} \rho_{\rm n} w^2 + \frac{1}{\gamma - 1}\frac{k_{\rm B}}{m_{\rm H}}\alpha \rho_{\rm c} \rho_{\rm n} (T_{\rm c} - T_{\rm n}) \label{eq:Qn},\\
  Q_{\rm c} = \frac{1}{2}\alpha \rho_{\rm c} \rho_{\rm n}w^2 - \frac{1}{\gamma - 1}\frac{k_{\rm B}}{m_{\rm H}}\alpha\rho_{\rm c} \rho_{\rm n} (T_{\rm c} - T_{\rm n}). \label{eq:Qc} 
\end{gather}
\end{subequations}
The first term in Eqs \eqref{eq:Qn} and \eqref{eq:Qc} is called the frictional heating (FH) \citep[see e.g.][]{DraRobDal1993aa, Dra1986aa,LeaDeVTha2014aa,Ni_LukMur2018aa}. It depends on the squared modulus of the drift velocity between charges and neutrals. The second term is the thermal exchange (TE). While the FH is always positive leading to the same heating rate for both components, TE sign can vary and it is the opposite for neutrals and charges (implying that it tends to homogenise the temperature of both species). The sum of $Q_{\rm n}$ and $Q_{\rm c}$
\begin{equation}
    Q_{\rm n} + Q_{\rm c} = \alpha \rho_{\rm c} \rho_{\rm n} w^2 \label{eq:QnQc}
\end{equation}
accounts for the variation of the total internal energy due to the collisions, finding that it is strictly positive, as the TE terms cancel each other. Then, we can say that differences in velocity between neutrals and charges contribute to increase the total internal energy of the plasma and, therefore, its temperature, but temperature differences between charges and neutrals do not \citep[see, e.g., the discussion by ][]{Pop2020aa}.  

Regarding the 1F equations, from Eqs. \eqref{eq:mom} and \eqref{eq:eq_e}, it follows 
\begin{subequations}
\begin{gather} \frac{\partial e_{\rm int}}{\partial t} + \nabla \cdot (e_{\rm int} \vec{u}) = - p \nabla \cdot \vec{u} + Q, \label{eq:eq_e_in}\\
    Q = \eta_{\rm A} J_{\bot}^2 = \tilde{\eta}_{\rm A} (\nabla \cdot \hat{p}_{\rm m})^2, \label{eq:Qamb}
\end{gather}
\end{subequations}
where $\vec{J}_{\bot} = \vec{J} \cdot (\hat{I} - \vec{b}\otimes \vec{b})$ is the component of the current density perpendicular to the magnetic field, with $\vec{b} = \frac{\vec{B}}{B}$. Again, we find that $Q$ is strictly positive, indicating that the internal energy of the centre of mass will increase, similar to the total internal energy in the 2F approach.

\section{Vertical propagation of MA waves in a stratified
atmosphere with a homogeneous horizontal magnetic field. Numerical experiments \label{sec:simulations}}


\begin{figure}[!t]
\includegraphics[width=\hsize]{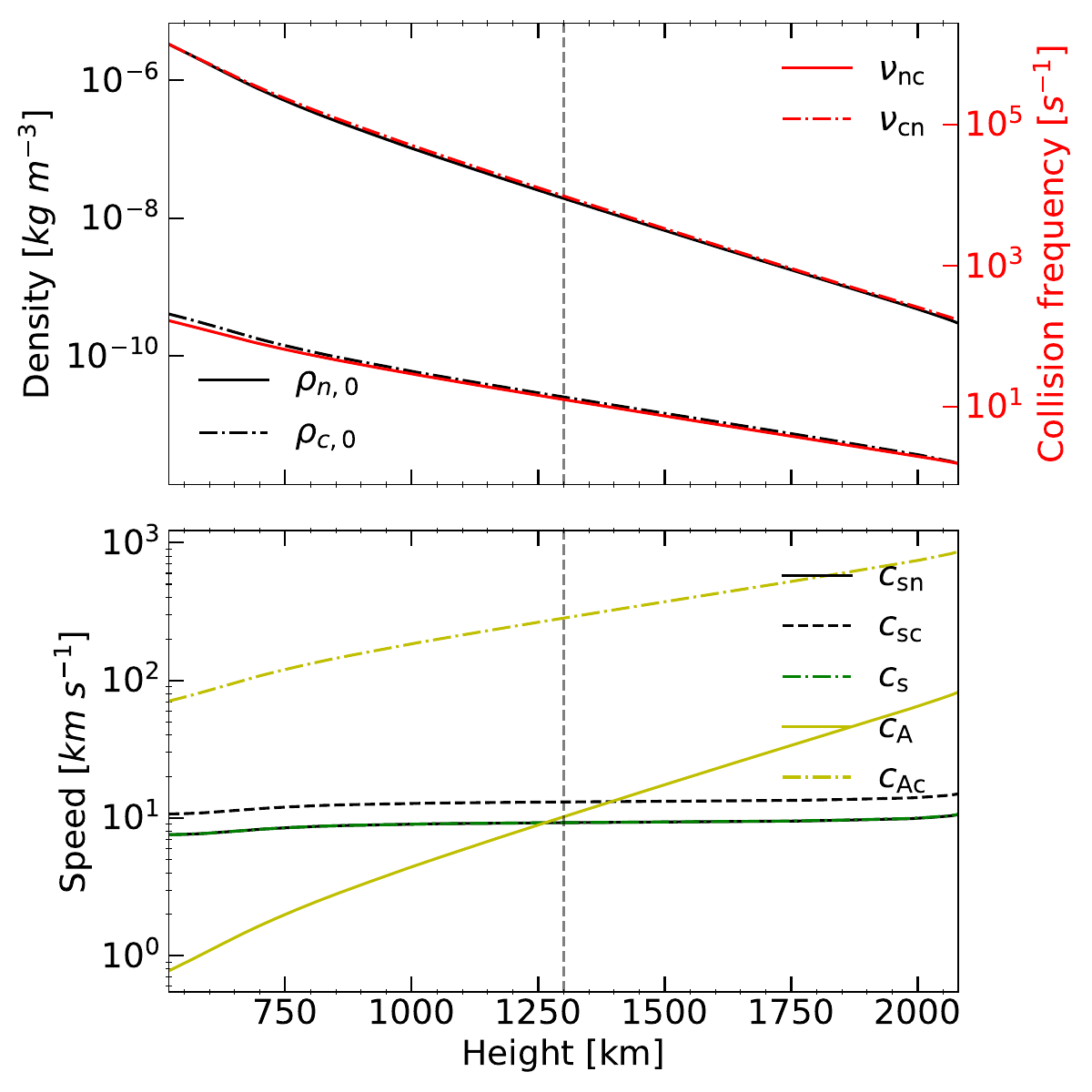}
\caption{Representation of the characteristic parameters of the atmosphere. In the top panel, the density of neutrals is in solid black line and the density of charges is in dashed-dotted line. Their scale is shown on the left axis. The right axis labels represent the neutral-charge (solid red line) and the charge-neutral (dashed-dotted red line) collision frequencies. The characteristic speeds are displayed in the bottom panel: the neutral (solid black line), charge (dashed line) and 1F (dashed-dotted green line) sound speeds and the 1F (solid yellow line) and charge (`dashed-dot' yellow line) Alfvén speeds. The vertical dashed line indicates the equipartition layer $\beta_\text{plasma} = 1$.}
\label{fig:equilibrium}
\end{figure}


\begin{figure*}[!t]
\centering
\begin{minipage}{0.68\textwidth}
\begin{subfigure}{.935\textwidth}
\includegraphics[width=\textwidth]{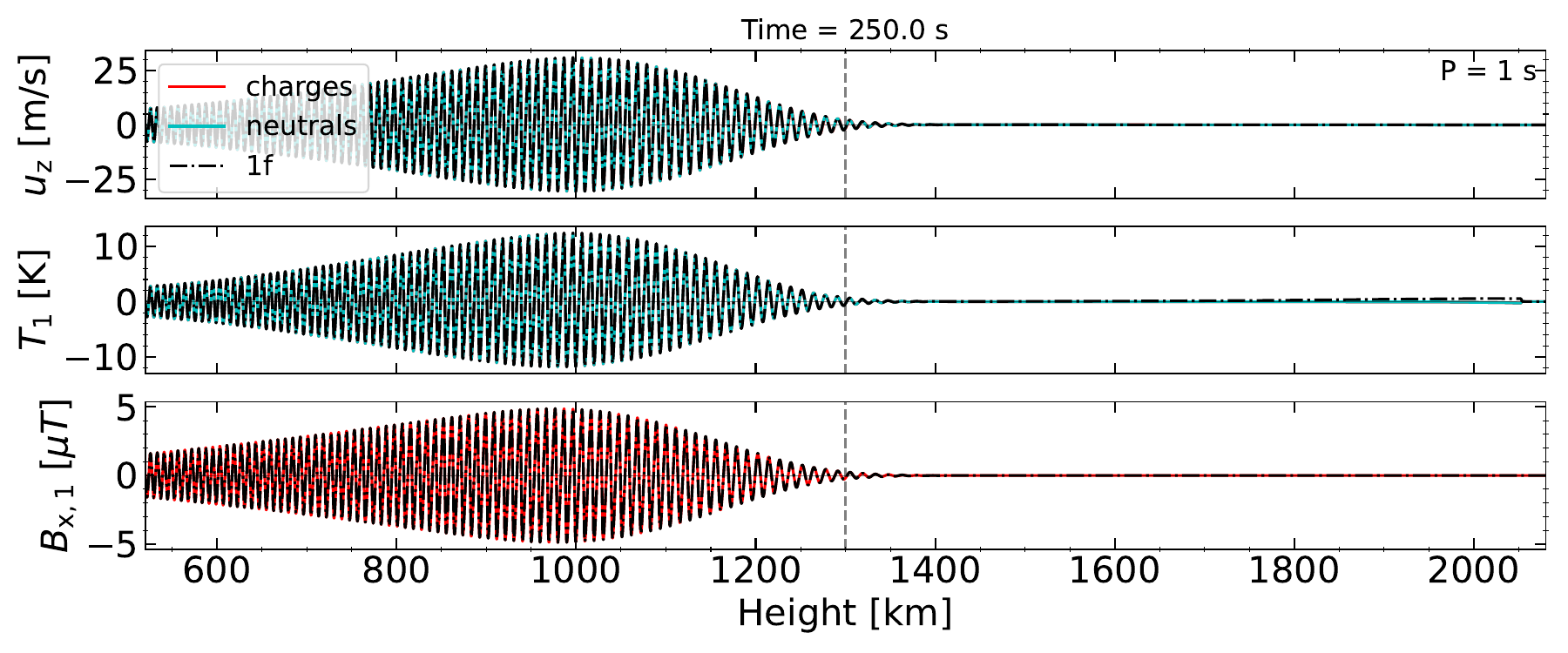}
\end{subfigure}
\begin{subfigure}{.935\textwidth}
\includegraphics[width=\textwidth]{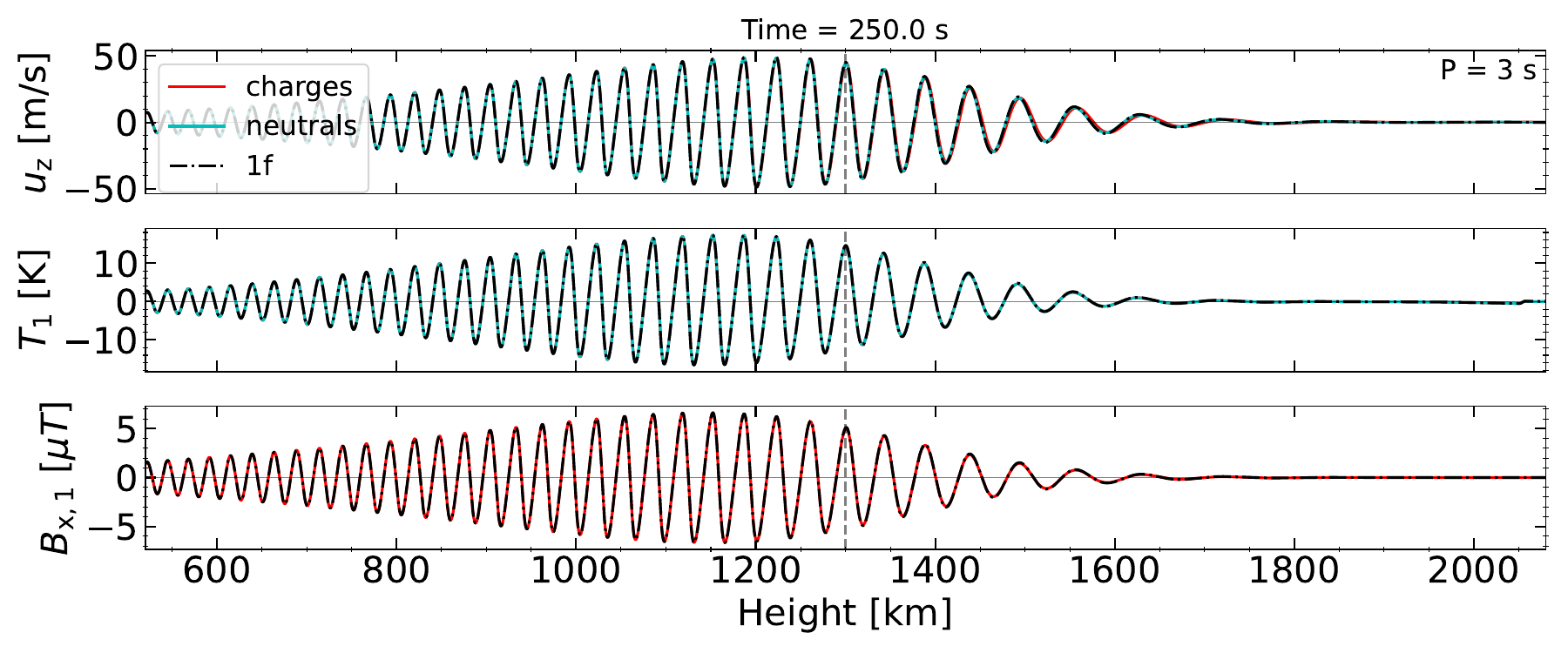}
\end{subfigure}
\begin{subfigure}[b]{.935\textwidth}
\includegraphics[width=\textwidth]{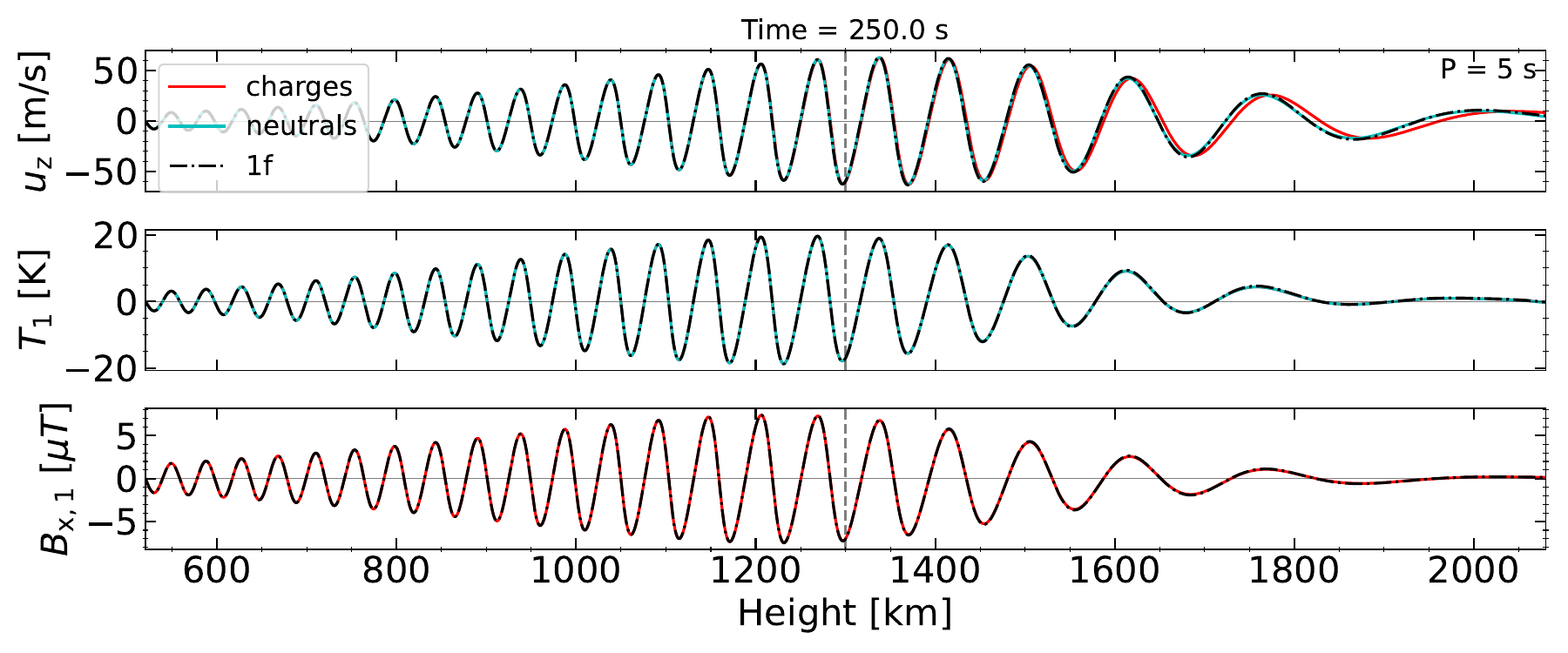}
\end{subfigure}
  \end{minipage}%
\begin{minipage}{0.30\textwidth}
\caption{Groups of panels from top to bottom: comparison between the numerical solution of velocity, temperature and magnetic field perturbations for waves of period $P = 1$s, 3s and 5s at 250 s after the start of the simulation. Solid red lines: perturbations in the velocity, temperature of charges and perturbations in the magnetic field in the 2F model; solid cyan lines: same for neutrals in 2F model; black dashed lines: same for the 1F model. The equipartition layer $\beta_{\rm plasma} = 1$ is indicated with a grey dashed vertical line.}
\label{fig:waves}
\end{minipage}
\end{figure*}

In this work, we solve both 2F and 1F equations using the code MANCHA-2F \citep{PopLukKho2019aa} based on the code MANCHA3D \citep{KhoCol2006aa, ModKhoVit2024aa}. Collision times vary by orders of magnitude through the chromosphere and in the deep layers they are much smaller than the typical time scales of hydrodynamic processes. Short collision times require small integration time steps for an explicit numerical scheme to satisfy the CFL condition. To handle this issue, MANCHA-2F introduces a semi-implicit scheme for updating the contribution of the collision terms in the 2F model, speeding up the computational time of the simulations. The code can also optionally solve the linearised system of equations, offering the possibility of separately assessing the linear and non-linear effects.

The numerical experiments are performed in a stratified atmosphere along the vertical axis ($z$-axis). The base is located at the bottom of the chromosphere, about 500 km above the solar surface. In this region, hydrogen is the primary donor of free electrons, which allows for a relatively self-consistent evaluation of charge density in our model. The top boundary is at the base of the transition region, where temperature starts to grow strongly. 

It must be noted that the approximations underlying the 2F model require that the  electron-proton collisional frequency is above the proton-neutral collisional frequency, the condition which is only fulfilled, on average, starting from the upper photosphere or the low chromosphere upwards \citep[see Figure 1 in][]{KhoColDia2014aa}, as well as \citet{VraKrs2013aa, SolBalZaq2015aa, BalAleCol2018aa}. In our work we consider heights starting from 600 km upwards, where this condition is mostly verified. Additionally, for the low wave frequencies considered in our work, it will be shown that both 1F and 2F models practically converge to the ideal MHD at the bottom part of the simulation domain \citep{Pid1956aa}. The application of both 1F and 2F models throughout the whole atmosphere is needed for consistency of their comparison. The ion-electron decoupling effects are not included in the 1F model either, so this aspect is not a source of discrepancies for the present discussion.

We set the equipartition region (where $\beta_{\rm plasma} = p / p_{\rm m} = 1$) at the middle of the physical domain. The background magnetic field is homogeneous and is oriented along the $x$-axis. We take the value of the background magnetic field to be $\sim 17$ G to set the equipartition region at the middle of the domain. The hydrostatic equilibrium is computed from a VAL3C temperature profile \mbox{\citep{VerAvrLoe1981}}, taking the mass density of hydrogen and electrons at the base of the chromosphere. This configuration resembles the one by \citet{PopLukKho2019ab}, except that the magnetic field is considered constant with height in the present work. Using the model with horizontal magnetic field allows reducing the problem to 1D propagation while still considering mode conversion, since the waves will change their nature from acoustic to magnetic on their way upwards. In the Sun, horizontal magnetic fields are frequently found in the canopy regions surrounding intergranular or supergranular magnetic field concentrations \citep{WieThaSol2014aa}. 

We impose a hydrostatic equilibrium (HS), in which each fluid is balanced by its own forces. As the background magnetic field is homogeneous, it does not contribute to the force balance. Then, pressure profiles are given by,
\begin{subequations}
\begin{gather}
    p_{\alpha,0}(z) = p_{\alpha,0}(0)\exp\left(-\int^z_0 \frac{dz'}{H_\alpha(z')}\right), \label{eq:palpha}\\[4pt]
    H_{\rm n} = \frac{k_{\rm B} T_{\rm n,0}}{m_{\rm H} g};~~~H_{\rm c} = \frac{2 k_{\rm B} T_{\rm c,0}}{m_{\rm H} g};~~~H = \frac{k_{\rm B} T_0}{\tilde{\mu}_0 m_{\rm H} g},
    \label{eq:Halpha}
\end{gather}
\end{subequations}
\noindent where $H_\alpha$ is the pressure scale height and the label `0' refers to the model atmosphere. We omit the index $\alpha$ for the 1F variables.  For a uniform magnetic field, $\vec{w}_0 = 0$, making the equilibrium states equivalent for both 2F and 1F models. However, if we consider the case of a non force-free magnetic field, i.e. $\nabla \cdot \hat{p}_{\rm m} \neq 0$, then $\vec{w}_0 \neq 0$ \citep[see discussion by][]{GoMMarKho2024aa}.

The characteristic speeds of the atmosphere are the sound speeds $c_{\rm{s\alpha}}$ and the Alfvén speeds $c_{\rm Ac}$ and $c_{\rm A}$, given by
\begin{gather}
    c_{\rm s} = \sqrt{ \frac{\gamma p_{0}}{\rho_{0}}};~~    c_{\rm s\alpha} = \sqrt{ \frac{\gamma p_{\alpha,0}}{\rho_{\alpha,0}}}; ~~c_{\rm A} = \frac{B_{\rm x,0}}{\sqrt{\mu_0 \rho_0}}; ~~c_{\rm Ac} = \frac{B_{\rm x,0}}{\sqrt{\mu_0 \rho_{\rm c,0}}} .\label{eq:calpha}
\end{gather}

The speeds $c_{\rm s}$ and $c_{\rm A}$ are representative for the highly coupled plasma, whereas $c_{\rm s\alpha}$ and $c_{\rm Ac}$ apply to the neutral and charged fluids when they are non-coupled. Here, we focus on the first case. 

Fig. \ref{fig:equilibrium} illustrates the atmospheric variables. The collision frequencies mainly depend on the density, decreasing exponentially with height. The hydrostatic equilibrium determines the ionisation fraction, which takes into account only the initial conditions at the base and does not include ionisation/recombination effects. The ionisation fraction (which coincides with the charge mass fraction) varies with height due to the different pressure scales of neutrals and charges from $10^{-4}$ at the bottom of our atmosphere to $10^{-2}$ at its top. Due to the low ionisation fraction in this model atmosphere, 1F profiles of thermodynamic variables and characteristic speeds closely resemble those for neutrals. The sound speeds are nearly constant, reflecting the fact that the temperature varies smoothly. The Alfvén speeds exhibit the most significant differences due to the low number of charges. However, if the components of the plasma are highly coupled, only the 1F Alfvén and sound speeds are relevant for the study of waves \citep{Pid1956aa, DePHae1998aa, KumRob2003aa, SolCarBal2013aa, SolCarBal2013ab}. 

For the numerical experiments, we launch a vertically propagating fast magneto-acoustic wave of a certain period from the bottom of the atmosphere. With MANCHA-2F, we solve both linear and non-linear systems of equations. Unless specified otherwise, we will discuss the non-linear solutions. The driver introduces the pure fast magneto-acoustic mode as a perturbation by fixing the amplitude of $u_{\rm c,z}$ to be $10^{-3}c_{\rm s}$, taking the local value of the sound speed at the bottom layer. The other variables are then evaluated according to the analytical solution. Details regarding the driver configuration can be found in Appendix \ref{App:A}. At the bottom, charges and neutrals are highly coupled, which means they move at the same velocity, $u_{\rm z}$ \citep{ZaqKhoRuc2011aa, SolCarBal2013ab}. Therefore, we set the amplitude of $u_{\rm z}$ in the 1F driver to be the same as that in the 2F driver. We perform a set of 2F and 1F simulations, varying the wave-period and we select the cases of periods P = 1, 3 and 5 s, whose frequency is close to the collision frequency at the high layers of the atmosphere (about 2 $s^{-1}$). 

The understanding of the numerical effects in the simulations is key for this analysis, especially since 1F and 2F employ different numerical schemes (explicit vs semi-implicit). In favour of performing a fair comparison, we restrict ourselves to the study of low amplitude waves, despite the amount of heating we produce is negligible when compared to the background temperature. Low amplitudes allow for low non-linear effects, and we do not need to include any hyper-diffusivity or shock diffusivity for controlling numerical effects. We fix the time step to be the same for 1F and 2F simulations, choosing in all the cases the most restrictive one. We also perform tests for different spatial resolutions, finding that for the cases of 5 s and 3 s periods, $dz \approx 500 \text{ m}$ is enough, but the 1 s period requires 4 times finer resolution. For the numerical stability reasons, a 6th order low-pass filter is applied to the solution each certain amount of simulation time, which is the same for 1F and 2F simulations. In all the experiments, we set a Perfectly Matched Layer (PML) \citep{Ber1994aa, ParKos2007aa} of the same size and parameters, at the top boundary. In addition to these numerical configurations, we also study the effects of the numerical scheme in the solutions. The 1F code uses a 4th order Runge-Kutta method for the integration, while the 2F code uses a semi-implicit method for the collisional terms, which is at most second order \citep[for more details, see][]{PopLukKho2019aa}. The 2F code has the option for computing the solution using an explicit scheme, the same as the 1F code. We have performed 2F simulations with both schemes, finding no significant differences between the solutions for our resolutions. Based on all of these arguments, we are confident that the differences between the solutions are caused by physical differences between the models.

The results of the simulations are shown in Fig. \ref{fig:waves}. At the bottom of the atmosphere, the gas pressure dominates over magnetic forces, and the waves behave essentially like acoustic-gravity waves. Since the waves move from more dense layers to less dense layers, the amplitude of velocity perturbation grows along the wave path. At a certain height, wave dissipation due to charge-neutral collisions becomes sufficiently effective in overcoming growth due to stratification and the amplitude of the waves starts to decay. As noted by \citet{PopLukKho2019aa}, \citet{PopLukKho2019ab} or \citet{ZhaPoeLan2021aa}, we also observe that damping reduces the non-linear effects of the wavefront. As expected, collisional damping is more effective for short-period waves, since their frequencies are closer to the collision frequencies. 

Broadly speaking, the 1F fluid solution agrees with the 2F results. As shown by \citet{PopLukKho2019ab}, the 1F solution reproduces essentially the neutral fluid's behaviour, since neutrals dominate in number. In line with their findings, we also observe a change in phase between charges and neutrals at the upper layers, leading to a velocity drift between the species. This difference is significant for the velocities, while temperatures remain similar for charges and neutrals because of the TE. Magnetic field oscillations are also very close for both models.

\begin{figure}[!t]
   \centering
   \includegraphics[width=\hsize]{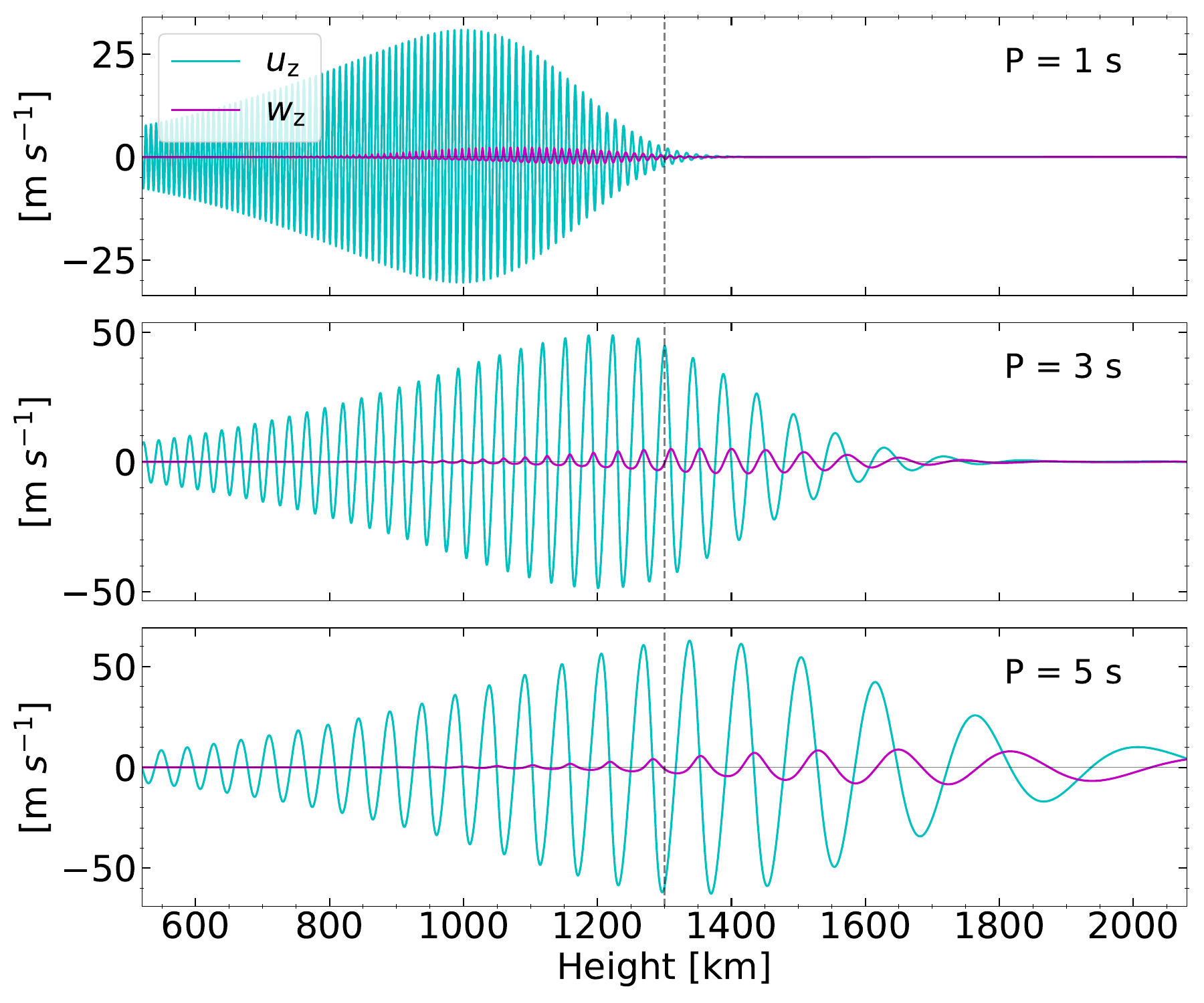}
      \caption{Comparison of the centre of mass velocity ($u$, cyan curves) and drift velocity ($w$, magenta curves) for the simulations represented in Fig. \ref{fig:waves}, at 250 s of simulated time. The values were computed from the 2F simulations. The vertical dashed line indicates the equipartition layer $\beta_\text{plasma} = 1$.}
         \label{fig:uvsw}
\end{figure}

When analysing the solutions depending on the wave-period, it leads to the conclusion that long-period waves exhibit greater decoupling effects, which may seem surprising. More in detail, we represent in Fig. \ref{fig:uvsw} the centre of mass velocity and the drift velocity from the 2F simulations, computed as Eqs. \eqref{eq:u_CM} and \eqref{eq:w}. We find $u \gg w$ in most of the domain, except for the layers where the wave is more damped, finding there that $u \approx w$. Due to the strong wave damping, it is less evident for a 1 s wave-period. However, a zoom-in inspection of the oscillations above the equipartition region reveals $u \approx w$. As stated before, the wave of larger period is the one leading to the largest drift velocity.

The 1F model is a valid description of plasma phenomena with characteristic timescales exceeding those of collisions. This model assumes that plasma components rapidly adjust their momentum and energy, behaving as a single fluid. When the timescales become closer to, but are still larger than collision times, ambipolar diffusion and other non-ideal effects due to neutrals may become relevant to describe the plasma behaviour in the 1F model. This limit was explored for 2F waves by \citet{ZaqKhoRuc2011aa}, \citet{SolCarBal2013aa}, and \citet{SolCarBal2013ab}, who performed a normal mode analysis for a homogeneous atmosphere, and more recently \citet{PopLukKho2019ab} extended the study to a vertically stratified atmosphere. Our numerical experiments show that increasing the wave-period in the range of periods we have explored leads to larger values of the drift velocity at the upper layers. This was already shown by other authors as \citet{PopLukKho2019ab}, where they compared the standard deviation of $w_{\rm z}$ for different wave-periods.
This result can be clarified by noting that shorter-period waves are more effectively damped, preventing them from reaching the upper layers where the most notable variations between the two models are expected. Conversely, large period waves can reach upper layers, where the fluids are less coupled. Therefore, it is not only important to check for the ratio $\omega / \nu_{\rm nc}$ when studying decoupling effects, but also consider the efficiency of the wave for transporting energy to layers of low coupling. If we continue increasing the wave-period, the trend must change because both the 2F and 1F models must converge to the ideal MHD model \citep[see, e.g.,][]{ZaqKhoRuc2011aa,SolCarBal2013aa,SolCarBal2013ab,CalGoM2023aa}. We do not explore this question in this work, but it can be anticipated from Fig. 6 by \citet{PopLukKho2019ab}, who studied the propagation of waves of periods from 1 s up to 20 s, finding that waves about 5 s period are the ones producing the largest drift velocities.

Several works, namely, \citet{Bra1965aa}, \citet{ZaqKhoRuc2011aa}, or \citet{KhoColDia2014aa} have derived the set of 1F equations from multifluids, obtaining terms proportional to $\vec{w}_\alpha$. Equations \eqref{eq:wn} and \eqref{eq:wc} show that for the case of a pure hydrogen plasma, $\vec{w}_\alpha \propto \vec{w}$ (it also stands if there are more components and all the neutrals move at velocity $\vec{u}_{\rm n}$ and all the charges at velocity $\vec{u}_{\rm c}$). We also have that for our weakly ionised model atmosphere, $\vec{w}_{\rm n} \rightarrow 0$ and $\vec{w}_{\rm c} \approx \vec{w}$ (see Eqs. \ref{eq:wn} and \ref{eq:wc}). Fig. \ref{fig:uvsw} shows that $\vec{u}$ and $\vec{w}$ become comparable as the waves reach upper layers, and discrepancies between 1F and 2F models could appear. We will discuss this point in more detail in section \ref{sec:discrep}.

\begin{figure*}[!t]
\centering
\includegraphics[width=\hsize]{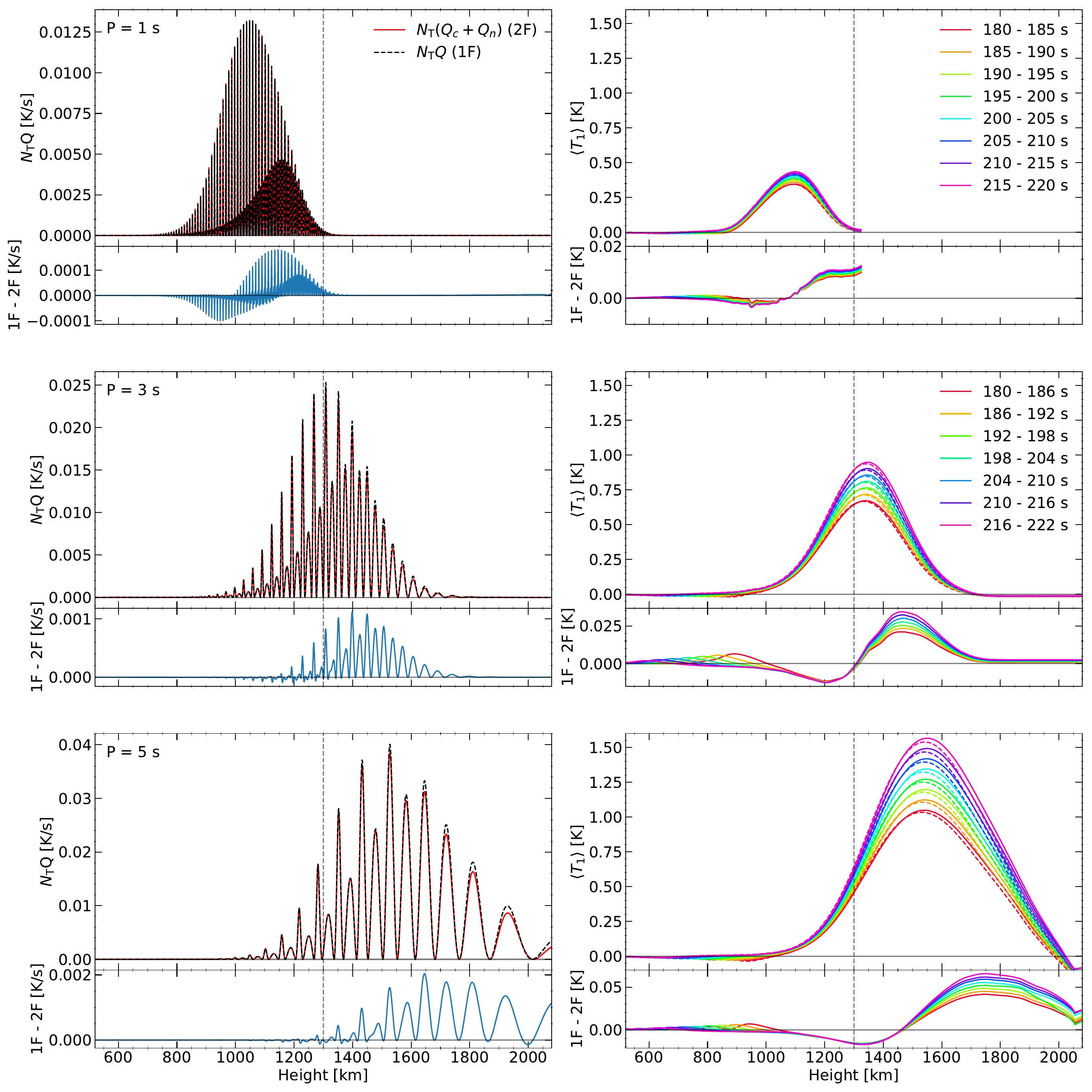}
\caption{Left column: sum of the 2F collisional heating terms (solid red line, see Eqs. \eqref{eq:Qc} and \eqref{eq:Qn}) and the 1F ambipolar heating term (black dashed line, see Eq. \eqref{eq:Qamb})), at 190 s of simulated time. The factor $N_{\rm T}$ (see Eq. \ref{eq:NT}) is introduced in all the cases. Right column: time-average temperature for the 1F and 2F models (solid and dashed lines, respectively). Different time intervals are represented with different colours. The vertical axis range is the same in all the panels. The 2F curves are computed from Eq. \eqref{eq:T_1F}. From the top to the bottom, the rows refer to the 1s, 3s, and 5s wave-periods. The averaging time intervals are the same for the cases of 1 s and 5 s. We also include residuals (1F - 2F) to better compare the differences between the models. The vertical dashed line indicates the equipartition layer $\beta_\text{plasma} = 1$.}
\label{fig:heating}
\end{figure*}

\begin{figure}[!t]
   \includegraphics[width=\hsize]{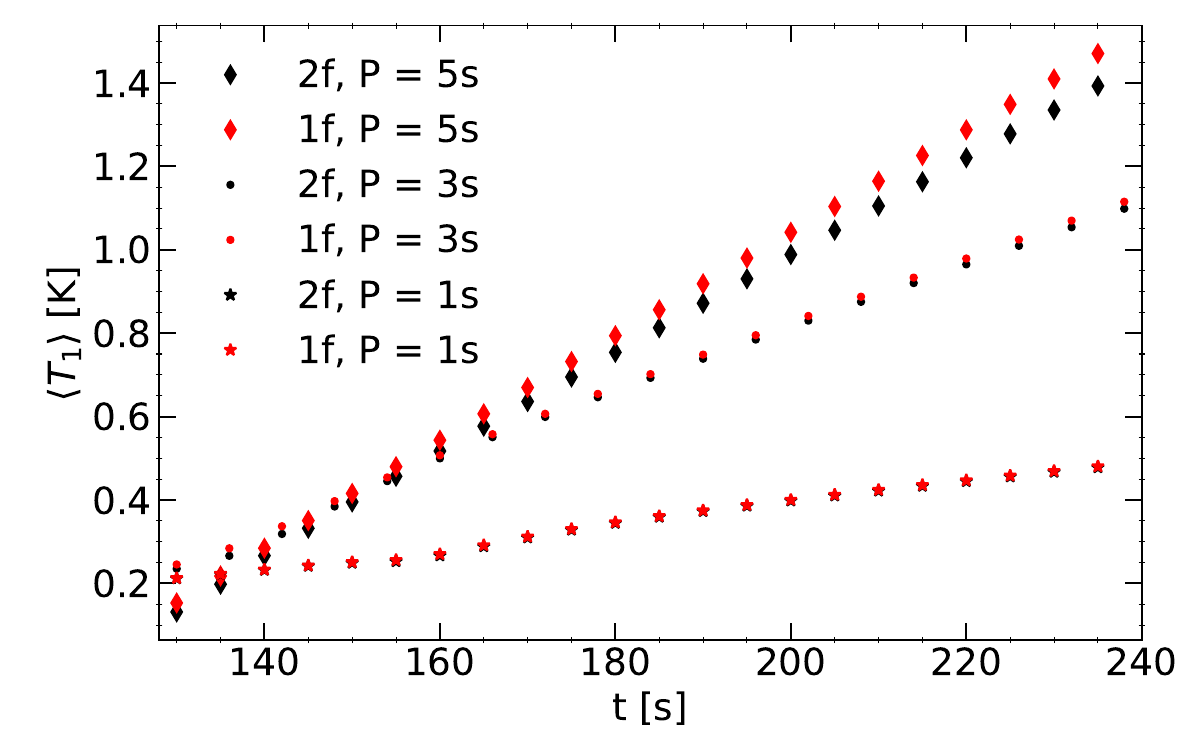}
      \caption{Evolution of the average temperature increase at the heating peak as a function of time. Diamonds: $P = 5$ s; dots: $P = 3$ s; stars: $P = 1$ s. The 2F data points are represented in black and the 1F ones in red. 
         \label{fig:Tvst}}
\end{figure}

\begin{figure}[!t]
   \centering
   \includegraphics[width=\hsize]{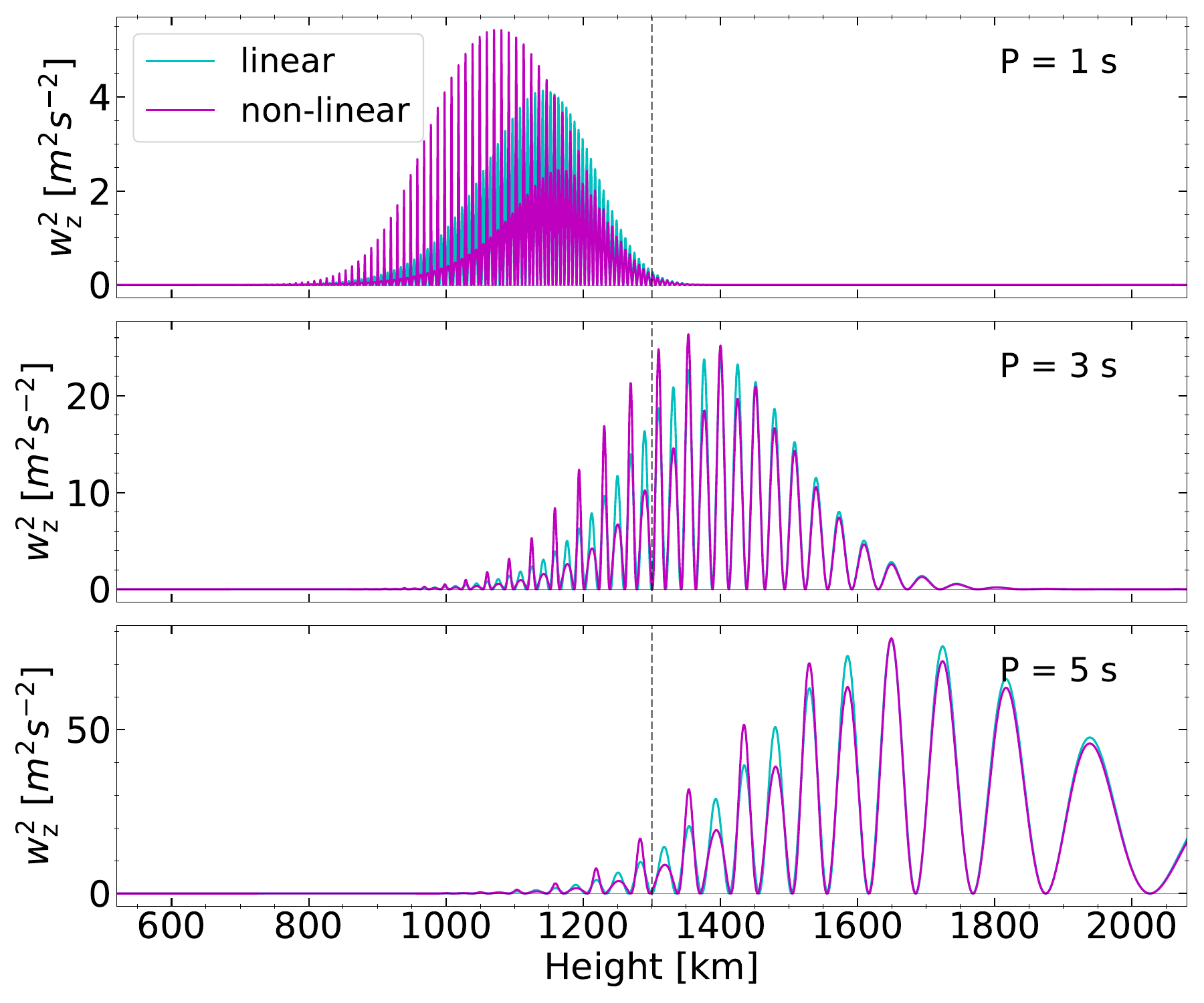}
      \caption{Comparison of $w_{\rm z}^2$ for a linear (cyan line) and non-linear (magenta line) 2F simulations, at 250 s of simulated time. From the top to the bottom, the wave-periods are P = 1, 3 and 5 s, respectively. The vertical dashed line indicates the equipartition layer $\beta_\text{plasma} = 1$.}
         \label{fig:wvsw}
\end{figure}

One of our primary objectives is to compare the heating effects in the 2F and 1F models. In this context, we represent in Fig. \ref{fig:heating} (left column) the source terms resulting from collisional and ambipolar heating of each model, as defined by Eqs \eqref{eq:Qn}, \eqref{eq:Qc}, and \eqref{eq:Qamb}, respectively. We multiply them by the factor
\begin{equation}
    N_{\rm T} = \frac{(\gamma - 1) m_{\rm H} \tilde{\mu}}{\rho k_{\rm B}}, \label{eq:NT}
\end{equation}
following the work by \citet{PopLukKho2019ab}, so they are expressed in $K~s^{-1}$.
At the bottom of the domain, the plasma is highly coupled and the heating terms play little role, and the wave behaves as expected from the ideal MHD model. As the waves propagate upwards, the collisional damping becomes more and more significant and the amplitude of the heating terms grows up until they reach a maximum. The peak position is higher for longer periods, consistently with the view of \citet{Ost1961aa}. When comparing these terms with the wave behaviour shown in Fig. \ref{fig:waves}, we observe that the position of the heating peaks are located after the velocity amplitude starts to be damped. The wave amplitude continues decaying after the peak, and the wave has less energy to release. Consequently, the increment of temperature decays as well. Most of the differences between the 2F and 1F solutions are observed in the upper layers for longer-period waves (see left panels in Fig. \ref{fig:heating}). For the 1 s period wave, we observe a transition in the behavior of heating rates. In deep layers, 2F heating has a higher amplitude than the 1F one and moving up, the 1F heating is dominating. Ambipolar heating is driven by magnetic forces, and the good agreement between model results proves that ambipolar diffusion is the main contributor to the drift velocity, hence, the impact of the $\vec{G}$ term (see Eq. \eqref{eq:w_1F}) on the solution for the atmosphere tested here is negligible. Furthermore, magnetic forces are able to drive larger drift velocities between neutrals and charges than pressure forces, as consequence of the low ionisation fraction of the atmosphere \citep{KhoArbRuc2004aa,KhoRucOli2006aa,PinGalBac2008aa}.

Focusing on the key variable, the temperature, we compute the average temperature increase by averaging the temperature perturbation over a wave-period for each layer of the atmosphere. For the case of the 1-second wave period, we chose to average several periods to reduce the noise. We search for the peaks and valleys of the wave and construct the wave envelopes by interpolation. Then, we compute the mean value of the top and bottom envelopes and, finally, the time average. For the 2F wave, we compute the centre of mass temperature, so temperatures of both models can be compared directly. By combining Eqs. \eqref{eq:p} and \eqref{eq:p_t}, it follows,
\begin{equation}
    T = \frac{\tilde{\mu} m_{\rm H}}{\rho k_{\rm B}} \sum_{\{n,c\}} \left(p_\alpha + \frac{1}{3} \rho_\alpha w_\alpha^2\right) = \frac{\tilde{\mu} m_{\rm H}}{\rho k_{\rm B}} \sum_{\{n,c\}} p_{\alpha} 
    + \frac{1}{3}\frac{\tilde{\mu}m_{\rm H}}{k_{\rm B}}\xi_{\rm c} \xi_{\rm n} w^2,
    \label{eq:Drift_T}
\end{equation}
where we apply Eqs. \eqref{eq:wn} and \eqref{eq:wc} on the last step.

We can safely neglect the second term in Eq. \eqref{eq:Drift_T} because $\xi_{\rm c} \ll 1$ and $\rho w^2 \ll \sum p_\alpha$, being null at the equilbrium. Then, we find
\begin{equation}
    T \approx \tilde{\mu} (\xi_{\rm n} T_{\rm n} + 2 \xi_{\rm c} T_{\rm c}). \label{eq:T_1F}
\end{equation}
This expression is similar to the one used, for example, by \citet{PopLukKho2019ab}. Those converge for a low ionisation fraction, but may lead to different values when the plasma is highly ionised. 

The comparison of the centre of mass temperatures from each model is shown in the right panels of Fig. \ref{fig:heating}, for different time ranges. It is apparent that the average temperature increases over time in both models. For all the wave-periods, differences between the 2F and 1F models are minor, but they become more evident at heights above the peak in temperature increase. Below the peak, differences are subtle but still discernible.
After the peak, the 1F model gives a more significant average temperature increase than the 2F model. We observe the opposite behaviour at heights before reaching the $T_1$ peak. The same pattern applies for all the considered periods. We have excluded the upper layers from the calculation for the 1-second period wave because the wave amplitude is too small and the computation becomes dominated by noise.  It is not straightforward to compute the temperature rate per time for each layer from the heating terms and also the method for computing the average temperature may introduce some systematics. However, we can state that both the heating terms and the average temperature increase exhibit a similar behaviour but, quantitatively, we cannot confirm if the computations match.

Having examined how the temperature changes as a function of height in the atmosphere, we study next how the average temperature increase changes as a function of time. 
In Fig. \ref{fig:Tvst}, we present the average temperature increase at the heating peak (different for the three periods) over time for each of the simulations. The short-period wave data points overlap, indicating a good agreement between the 2F and 1F models. Conversely, 
there are more differences visible for the longer-period waves. Since the wave energy is proportional to the wave amplitude squared, if there is more energy available, then it is possible to increase more the internal energy of the plasma, even if the dissipation is less effective. These results are consistent with Fig. 14 by \citet{ZhaPoeLan2021aa}, where they represented $\langle T_1 \rangle$ for different wave-periods, but they were not including a magnetic field in their study. 

It is important to highlight the linear dependence of $\langle T_1 \rangle$ on time, in agreement with what was previously reported by \citet{PopLukKho2019ab} from their simulations with a similar setup. This linear dependence is consequence of launching waves of small amplitude, meaning that energy dissipation is solely dependent on the initial background state and the wave frequency \citep{Bra1965aa,  Pri2014aa,TraBriRic2014aa}. This increment is minor compared to the background temperature, however, one has to take into account that we use very low initial amplitudes, about 10 ${\rm m} \ {\rm s}^{-1}$. In the solar case, at the upper photosphere, the wave amplitudes reach $\sim 1000~ {\rm m} \ {\rm s}^{-1}$ \citep[see, for example,][]{BecKhoRez2009aa}, so $\langle T_1 \rangle$ will be proportionally larger. 

The analysis of the heating terms represented in Fig. \ref{fig:heating} brings another curious feature. We observe two groups of peaks whose envelopes have different amplitudes and shapes. This behaviour is more clear when looking at the case of 1 s wave-period. From Eq. \eqref{eq:Qn} and \eqref{eq:Qc}, it follows that $Q_{\rm n} + Q_{\rm c} \propto w_{\rm z}^2$. We show in Fig. \ref{fig:wvsw} the shape of $w_{\rm z}^2$ for all the wave-periods simulated, obtaining that $w_{\rm z}^2$ exhibits the same behaviour as $Q$. We have also computed $w_{\rm z}^2$ for the same setup, but solving the linear equations with MANCHA-2F. The comparison shows that the two groups are present in both cases, but in the linear simulation they follow the same envelope. Back to Fig. \ref{fig:uvsw}, if we look at one of the peaks of $u_{\rm z}$, the slope of the wave is steeper after the peak than before, for those more sharped. As consequence, the peaks and valleys in $w_{\rm z}$ become asymmetric due to the non-linear effects. Once the amplitude of $w_{\rm z}^2$ starts to decay, the non-linear effects become weaker and the amplitude of the two groups of peaks tend to the linear amplitude. It is interesting to observe that for the case of 1 s wave-period, the envelopes of $w_{\rm z}^2$ have the maximum values at different layers. A similar result was obtained by \citet{PopLukKho2019ab} (see Fig. 8) for the same setup. It is also possible to guess the same behaviour for the 3 s case. Despite the small amplitude of the waves, $w_{\rm z}$ is highly sensitive to the non-linear effects due to their high frequency.

\begin{figure*}[!t]
\centering
\begin{minipage}{0.68\textwidth}
\includegraphics[width=.935\hsize]{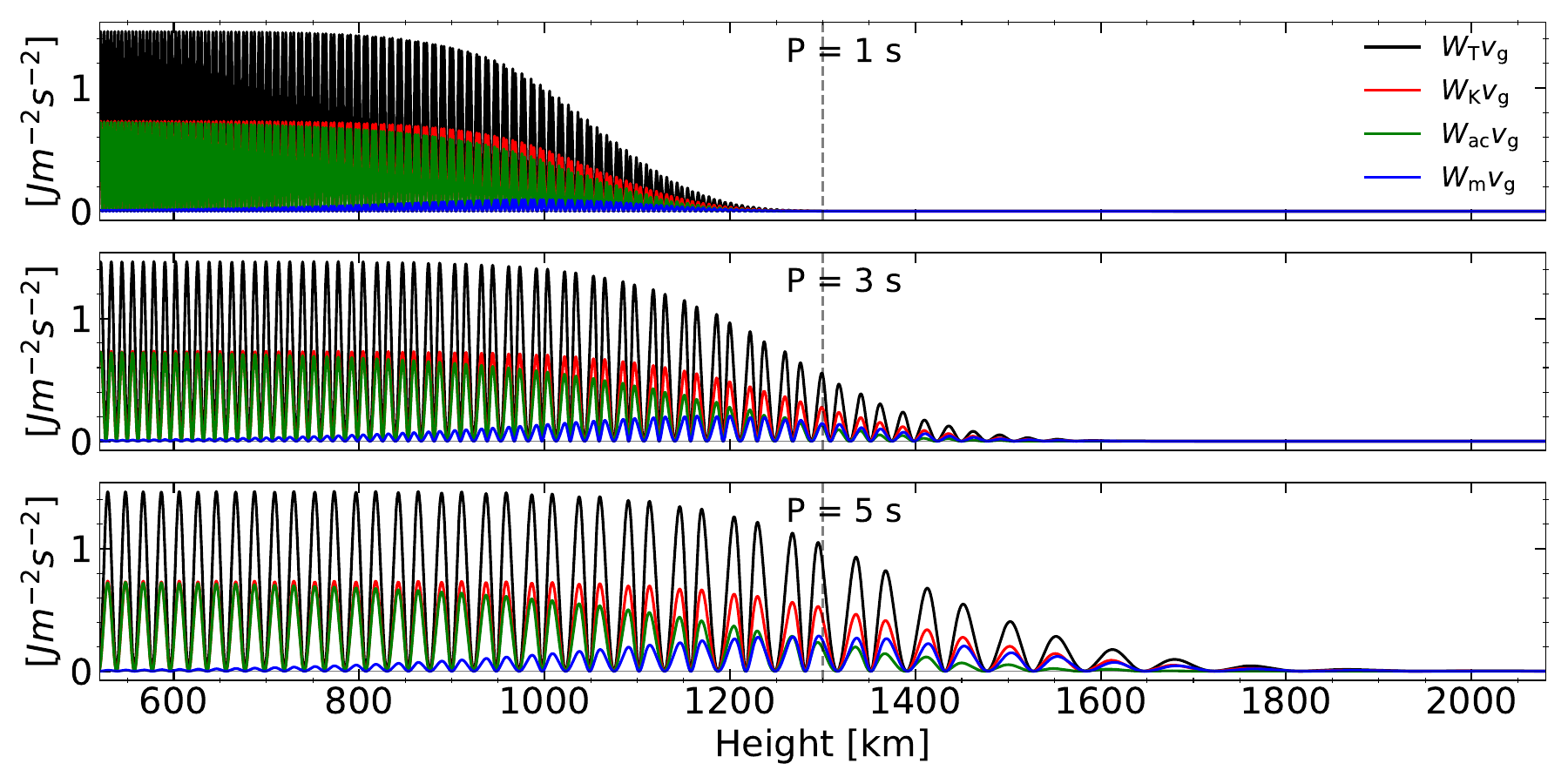}
\end{minipage}
\begin{minipage}{0.3\textwidth}
\caption{Representation of the 1F wave energy flux contributions for the three wave-periods studied in this work. From the top to the bottom, the wave-periods are 1 s, 3 s, and 5 s, respectively. Blue line: the magnetic energy ($W_{\rm{m}}$); green line: the acoustic energy ($W_{\rm{ac}}$); red line: the kinetic energy ($W_{\rm{K}}$), black line; the total energy ($W_{\rm{T}}$). The vertical dashed line indicates the equipartition layer $\beta_\text{plasma} = 1$.}
\label{fig:energy1F}
\end{minipage}
\end{figure*}

\begin{figure*}[!t]
\centering
\begin{minipage}{0.68\textwidth}
\includegraphics[width=.935\textwidth]{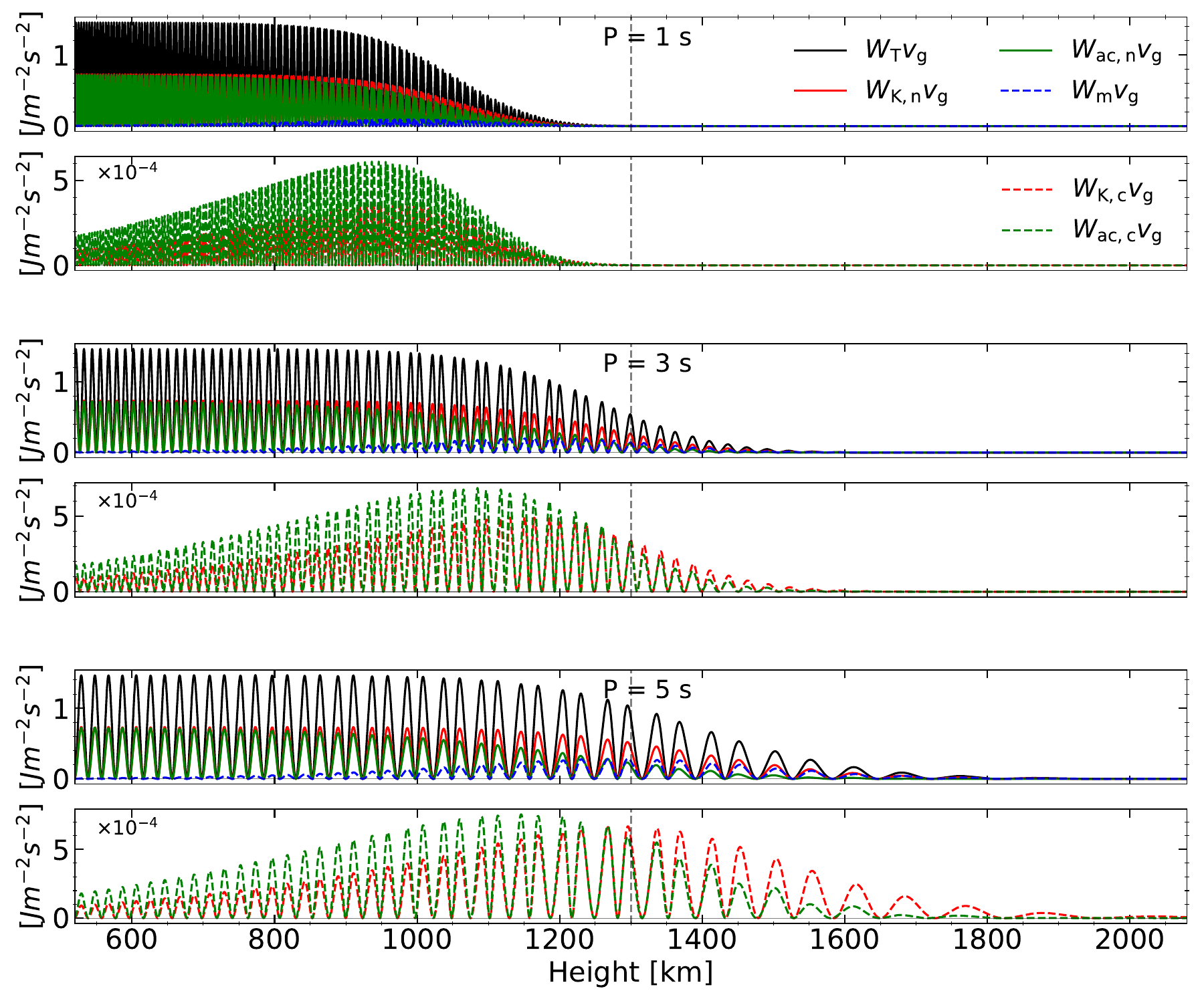}
\end{minipage}
\begin{minipage}{0.3\textwidth}
\caption{Representation of the 2F wave energy flux contributions for the three wave-periods studied in this work. From the top to the bottom  in groups of two plots, the wave-periods are 1 s, 3 s, and 5 s, respectively. Blue line: the magnetic energy ($W_{\rm m}$); green line: the acoustic energy ($W_{\rm ac,\alpha}$); red line: the kinetic energy ($W_{\rm K, \alpha}$), black line; the total energy ($W_{\rm T}$). The terms of the neutrals are shown in solid lines and terms for the charges in dashed lines. We split the terms $W_{\rm K,c}$ and $W_{\rm ac,c}$ in panels separated from neutrals due to the difference in scale. The vertical dashed line indicates the equipartition layer $\beta_\text{plasma} = 1$.}
\end{minipage}
\label{fig:energy2F}
\end{figure*}

We complete the analysis of the simulated waves by looking into the energy distribution. From \citet{Bra1965aa}, \citet{BraLou1974aa}, \citet{Wal2004aa}, \citet{Wal2014aa}, \citet{CalGoM2023aa} or \citet{Cal2023aa}, the wave energy terms can be defined as,
\begin{equation}
    W_{\rm K,\alpha} = \frac{1}{2}\rho_{\alpha,0}u_{\alpha}^2;~~~W_{\rm ac, \alpha} = \frac{p_{\alpha,1}^2}{2 \rho_{\alpha,0}c_{\rm s\alpha}^2};~~~W_{\rm m} = \frac{B_{1}^2}{2 \mu_0}.
    \label{eq:wave_energy}
\end{equation}
From the left to the right, they are the kinetic energy density, the acoustic energy density and the magnetic energy density. Compared with the works cited above, we do not include the buoyancy term here as we have $\omega \gg \nu_{\rm ac}$, where the acoustic cutoff frequency reaches a maximum value of $\nu_{\rm ac} = \frac{c_{\rm s}}{2 H}\approx 5$ mHz at the bottom of our atmosphere, which is located around the temperature minimum. Therefore, we consider total energy $W_{\rm T}$ to be only the sum of the kinetic, acoustic and magnetic energy, for all the fluid components considered in each model. Figs. \ref{fig:energy1F} and \ref{fig:energy2F} show the amount of energy of each type that the waves are carrying along the model atmosphere. Those terms are multiplied by the 1F group velocity of the fast mode, $v_{\rm g} = \sqrt{c_{\rm s}^2 + c_{\rm A}^2}$, so the total energy times the group velocity (the total wave energy flux) is constant in the 1F limit in absence of dissipation mechanisms, based on the wave-action conservation \citep[see, e.g.,][]{TraBriRic2014aa, Wal2014aa}. All the waves exhibit a constant energy flux at lower layers of the atmosphere, where the neutrals and charges are highly coupled by collisions. Once the collisional dissipation becomes effective, the total energy decreases. This was also noted by \citet{ZhaPoeLan2021aa} when analysing the acoustic energy of the waves. We also find that the distance from the layer where the wave starts to decay up to the layer where the wave effectively vanishes decreases for short wave-periods, which evidences the more efficiency of charge-neutral collisions in dissipation high frequency waves.  

When looking at the behaviour of the energy components in 1F (see Fig. \ref{fig:energy1F}), we observe that, at the lower layers, the wave energy is divided into kinetic and acoustic energy almost in the same proportion, with a small contribution of the magnetic energy. This behavior is the print of an acoustic wave, whose energy equally splits into acoustic and kinetic   \citep[see, e.g.,][]{LanLif1987aa,Wal2014aa}. As the wave propagates upwards, the kinetic energy keeps constant but the acoustic energy gradually exchanges to magnetic energy. At the point where energy dissipation becomes efficient, all the energies start to decay, but the exchange between acoustic to magnetic energy keeps the same trend. In the neighborhood of the equipartition region, the wave reaches a layer where acoustic and magnetic energies are in the same proportion, and as the wave continues its path, the acoustic energy is gradually exchanged into magnetic energy. In other words, the analysis of the wave energy shows that the wave transforms from acoustic to magnetic \citep[see, e.g., ][]{SchCal2006aa, KhoCol2006aa, Wal2014aa, CalGoM2023aa}. 

If we follow the same analysis for the 2F waves (see Fig. \ref{fig:energy2F}), we find that the energy terms related to the neutrals and magnetic energy show approximately the same behaviour as in the 1F waves. Regarding the kinetic and acoustic energy of charges, we find them to be orders of magnitude smaller than those of the neutrals because of the low ionisation fraction. In addition, at deep layers, those energy fluxes are clearly different; the acoustic energy approximately doubles the kinetic energy. This can be explained as follows. In those layers, the ideal MHD solution is still valid and the phase speed is essentially $c_{\rm s}$, therefore, $\omega \approx k_{\rm z} c_{\rm s}$. Then, from Eq. \eqref{eq:P}, it follows that $p_{\rm c,1} \approx \rho_{\rm c,0} c_{\rm s} u_{\rm c, z}$, where we have neglected the growth term since we are only interested in a small region and also assume that $u_{c,z} = u_{z}$, in light of Fig. \ref{fig:uvsw}. Introducing this relation into the expression for the acoustic energy for charges in \eqref{eq:wave_energy}, we obtain 
\begin{equation}
    W_{\rm ac,c} \approx \frac{ p_{\rm c,0}}{2 p_0} \rho_0 u_{c,z}^2 \approx \rho_{c,0} u_{c,z}^2, \label{eq:Wacdouble} 
\end{equation}
which is the double of the kinetic energy of the charges. In the last step, we have applied the limit of weakly ionisation. Unlike the case of the neutrals, $W_{\rm K,c}$ and $W_{\rm ac,c}$ grow up at the deep layers and then decay after peaking, much similar to the behavior of $W_{\rm m}$.

For all the wave-periods considered in this work, we observe that the dissipation becomes more efficient as more energy is exchanged into magnetic energy. \citet{ForOliBal2007aa} found the fast waves to be efficiently damped by charge-neutral collisions in prominences. In their study, the Alfvén speeds were above 10 km s$^{-1}$, while the temperature was 8000 K. For this temperature in the VAL3C model, the sound speed is less than 10 km s$^{-1}$, so their waves should have more magnetic than acoustic energy. By comparing Figs. 1 and 2 from \citet{SolCarBal2013ab}, we see that the wave dissipation is larger for $\beta_{\rm plasma} \ll 1$ regions than for $\beta_{\rm plasma} \gg 1$ regions. In addition, the works conducted by \citet{SolCarBal2015aa}, \citet{PopKep2021aa} or \citet{CalGoM2023aa} have shown that if a wave excites efficiently the magnetic field, such as in the case of fast and slow waves when they are magnetic (that is, when $\beta \ll 1$ and $\beta \gg 1$, respectively), or in the case of Alfvén waves, then charge-neutral collisions efficiently damps the wave.

\begin{figure}[!t]
\centering
\includegraphics[width=\hsize]{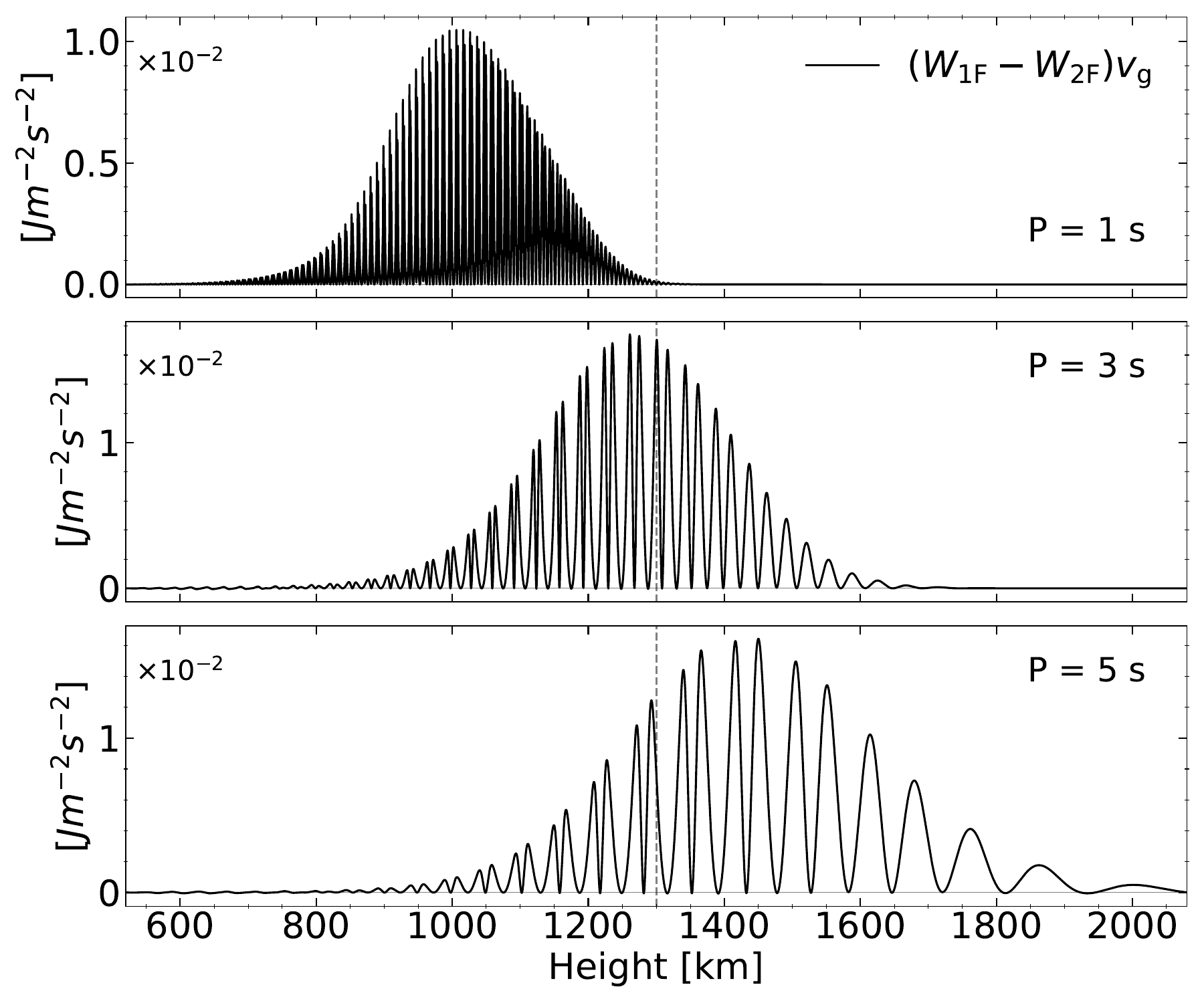}
\caption{Resudials of the total wave energy flux (1F - 2F). The results are computed for the different wave-periods considered in this work. The vertical dashed line indicates the equipartition layer $\beta_\text{plasma} = 1$.}
\label{fig:ediff}
\end{figure}

After we analyse the wave energy for each model, we compare the total wave energy from each one. In Fig. \ref{fig:ediff}, we show the difference in wave flux between the models. We observe that the envelope and peak position of the flux differences reminds the behaviour of the heating terms (see Fig. \ref{fig:heating}, or Fig. \ref{fig:wvsw}). We remark this fact, since it points to a physical origin for the total flux differences. Another important aspect is that the flux differences are mostly positive along the atmosphere, i.e., the wave in the 1F model is carrying more energy to the high layers than in the 2F model. The 2F model describes the dynamics of two fluids whereas the 1F model restricts itself to the centre of mass of the system, so one would expect the energy of the 2F model to be equal or overcome the 1F energy. Therefore, the result can be explained by considering that the 2F model is more efficient at dissipating energy at deep layers than the 1F model. This view also aligns with the temperature increase differences shown in Fig. \ref{fig:heating}, where we have noted that the 2F model increases the temperature more than the 1F model before the heating peak.

The analysis of the simulations brings us to some key points about wave heating through charge-neutral collisions. Commonly in literature \citep[namely, ][]{ZaqKhoRuc2011aa, SolCarBal2013aa, SolCarBal2013ab, KhoColDia2014aa} it is mentioned the ratio $\omega / \nu$, with $\nu$ the collision frequency of the considered charge-neutral model, as the criterion for considering the partially ionised effects. However, in line with the findings of \citet{PopLukKho2019ab} or \citet{ZhaPoeLan2021aa}, and this work, the most significant temperature increases are not driven by the short period waves. In those setups, the waves are launched at the bottom of the atmosphere and propagate vertically. Then, the long-period waves can transport more energy than short-period waves to the high layers, where the coupling between the species weakens. Therefore, the ratio $\omega / \nu$ is a good reference for a local analysis, but also the capacity of the wave for transporting energy should be taken into account. We also find that when the magnetic energy of the wave increases, the charge-neutral dissipation becomes more efficient, in agreement with the studies by \citet{ForOliBal2007aa}, \citet{SolCarBal2015aa}, \citet{PopKep2021aa}
or \citet{CalGoM2023aa}. Finally, we propose the efficiency of 2F model in dissipating energy at deep layers as the explanation for the discrepancy on the temperature increase. However, the physical assumptions that lead to a different dissipation between models are still not clear and will drive the discussion of the following sections.

\section{Vertical propagation of MA waves in a stratified
atmosphere with a horizontal and homogeneous magnetic field. Analytic approach \label{sec:eikonal}}

In light of the numerical experiments, we now apply an analytical approach to deepen into the solutions of each model and to better understand the discrepancies between the models. As we apply the same methodology for both models, we present below all the partial results in parallel. Our first step involves deriving the linear wave equations. We consider the case of vertical propagation (z-axis) in a stratified atmosphere, with a horizontal magnetic field in the x-axis. As usual, we split variables, say $f$, into equilibrium and perturbation parts $f = f_0 + f_1$, such as $|f_1|/f_0 \ll 1$, where the subindex ``0'' stands for the equilibrium and the subindex ``1'' for the perturbation. We consider a static equilibrium, so $u_{\rm \alpha,z,0} =0$ and $|\vec{u_{\alpha,z,1}}|$ much smaller than the phase speed. Henceforth, we remove the subindex ``1'' from the velocity perturbation. Keeping only the terms up to the first order, the set of 2F linear equations is \citep[see, e.g.,][]{PopLukKho2019ab},  
\begin{subequations}
\begin{gather}
\frac{\partial \rho_{\rm n,1}}{\partial t} + \frac{\partial}{\partial z }(\rho_{\rm n,0}u_{\rm n,z}) = 0, \label{eq:rhon_lin}\\
\frac{\partial \rho_{\rm c,1}}{\partial t} + \frac{\partial}{\partial z}(\rho_{\rm c,0}u_{\rm c,z}) = 0, \label{eq:rhoc_lin}\\
\rho_{n,0}\frac{\partial u_{\rm n,z}}{\partial t} + \frac{\partial p_{\rm n,1}}{\partial z} = - \rho_{\rm n,1} g + \alpha_0 \rho_{\rm c,0} \rho_{\rm n,0} (u_{\rm c,z} - u_{\rm n,z}), \label{eq:momn_lin}\\
\rho_{\rm c,0}\frac{\partial u_{\rm c,z}}{\partial t} + \frac{\partial p_{\rm c,1}}{\partial z} = - \rho_{\rm c,1} g - \alpha_0 \rho_{\rm c,0} \rho_{\rm n,0} (u_{\rm c,z} - u_{\rm n,z}) - \frac{B_{\rm x,0}}{\mu_0}\frac{\partial B_{\rm x,1}}{\partial z}, \label{eq:momc_lin}\\
\frac{\partial p_{\rm n,1}}{\partial t} + u_{\rm n,z} \frac{dp_{\rm n,0}}{dz} - c_{\rm sn}^2\left(\frac{\partial \rho_{\rm n,1}}{\partial t} + u_{\rm n,z} \frac{d \rho_{\rm n,0}}{dz}\right) = 0, \label{eq:en_lin}\\
\frac{\partial p_{\rm c,1}}{\partial t} + u_{\rm c,z} \frac{dp_{\rm c,0}}{dz} - c_{\rm sc}^2\left(\frac{\partial \rho_{\rm c,1}}{\partial t} + u_{\rm c,z} \frac{d \rho_{\rm c,0}}{dz}\right) = 0, \label{eq:ec_lin}\\
\frac{\partial B_{\rm x,1}}{\partial t} = - B_{\rm x,0} \frac{\partial u_{\rm c,z}}{\partial z}. \label{eq:induc2F_lin}
\end{gather}
\end{subequations}
Similarly, the 1F set of linear equations is
\begin{subequations}
\begin{gather}
    \frac{\partial \rho_1}{\partial t} + \frac{\partial}{\partial z}(\rho_0 u_{\rm z}) = 0, \label{eq:rho_lin} \\
\rho_0 \frac{\partial u_{\rm z}}{\partial t} + \frac{\partial p_1}{\partial z} =  - \rho_1 g - \frac{B_{\rm x,0}}{\mu_0}\frac{\partial B_{\rm x,1}}{\partial z}, \label{eq:mom_lin} \\
\frac{\partial p_1}{\partial t} + u_{\rm z} \frac{dp_0}{dz} - c_s^2\left(\frac{\partial \rho_1}{\partial t} + u_z \frac{d \rho_0}{dz}\right) = 0, \label{eq:e_lin} \\
\frac{\partial B_{\rm x,1}}{\partial t} = - B_{\rm x,0} \frac{\partial u_{\rm z}}{\partial z} + c_{\rm A}^2 \rho_0 \frac{\partial}{\partial z}\left(\tilde{\eta}_{\rm A,0} \frac{\partial B_{\rm x,1}}{\partial z}\right),\label{eq:induc1F_lin}
\end{gather}
\end{subequations}
where the only non-ideal contribution comes from the ambipolar diffusion in the induction equation.

By combining the linear equations, we obtain a set of two second order linear equations that involves the velocities of each fluid. For the 2F model, we find  
\begin{subequations}
\begin{gather}
    \frac{\partial^2 u_{\rm n,z}}{\partial t^2} = c_{\rm sn}^2 \frac{\partial^2 u_{\rm n,z}}{\partial z^2} + \frac{1}{\rho_{\rm n,0}}\frac{d}{dz}( c_{\rm sn}^2\rho_{n,0}) \frac{\partial u_{\rm n,z}}{\partial z}  \label{eq:wequnz}\\ + \alpha_0 \rho_{\rm c,0}\frac{\partial}{\partial t}(u_{\rm c,z} - u_{\rm n,z}), \nonumber\\
    \frac{\partial^2 u_{\rm c,z}}{\partial t^2} = (c_{\rm sc}^2 + c_{\rm Ac}^2) \frac{\partial^2 u_{\rm c,z}}{\partial z^2} + \frac{1}{\rho_{\rm c,0}}\frac{d}{dz}[(c_{\rm sc}^2 + c_{\rm Ac}^2)\rho_{\rm c,0}]\frac{\partial u_{\rm c,z}}{\partial z} \label{eq:wequcz}\\- \alpha_0 \rho_{\rm n,0}\frac{\partial }{\partial t}(u_{\rm c,z} - u_{\rm n,z}), \nonumber
\end{gather}
\end{subequations}
and, for the 1F model
\begin{subequations}
\begin{gather}
    \frac{\partial^2 u_{\rm z}}{\partial t^2} = c_{\rm s}^2 \frac{\partial^2 u_{\rm z}}{\partial z^2} + \frac{1}{\rho_0} \frac{d}{dz}(c_{\rm s}^2 \rho_0)\frac{\partial u_{\rm z}}{ \partial z}  + \frac{1}{\rho_0 \tilde{\eta}_{\rm A,0}}\frac{\partial}{\partial t}(u_{\rm Amb,z} - u_{\rm z}), \label{eq:wequz}\\
    0 = c_{\rm A}^2\frac{\partial^2 u_{\rm Amb,z}}{\partial z^2} - \frac{1}{ \rho_0 \tilde{\eta}_{\rm A,0}}\frac{\partial}{\partial t}(u_{\rm Amb,z} - u_{\rm z}),\label{eq:wequambz}
\end{gather}
\end{subequations}
where we introduce the charge velocity due to ambipolar diffusion, $u_{\rm Amb,z}$, as
\begin{equation}
    u_{\rm Amb,z} = u_{\rm z} - \tilde{\eta}_{\rm A,0}\frac{B_{\rm x,0}}{\mu_0}\frac{\partial B_{\rm x,1}}{\partial z}.\label{eq:uAmb}
\end{equation}
This definition follows from introducing $\vec{w}_{\rm 1F}$ (Eq. \eqref{eq:w_1F}) in relation \eqref{eq:uc_u}, only accounting for the magnetic term. In the 2F model, we find wave equations for the velocities of neutrals and charges. Conversely, a wave equation couples with a diffusion equation in the 1F model. We also note that a term proportional to $\frac{\partial u_{\rm Amb,z}}{\partial z}$, which accounts for the stratification, is absent from the equation of $u_{\rm Amb,z}$ because $B_{\rm x,0}$ is uniform.

Across this work, we have denoted the decay of the wave amplitude as damping or dissipation. In absence of other effects, wave dissipation implies that the wave amplitude decreases strictly because the wave is releasing energy to the environment. Note however that there is no dissipation term in the linear energy equations as that term is of the second order, therefore, heating cannot be properly captured on the linear approach \citep[see, e.g.,][]{CalGoM2023aa, Cal2023aa}. In the case of the stratified atmosphere, velocity waves increase their amplitude but not their energy flux due to the gravitational stratification. In terms of velocity amplitude, the growth by stratification compensates for the effects of the dissipation when dissipation is still weak. Consequently, Figs. \ref{fig:energy1F} and \ref{fig:energy2F} show that wave energy flux is constant at deep layers, and above them the energy flux drops down. 
The velocity waves we are studying propagate to layers with decreasing density, increasing their amplitude, and to layers of decreasing collision frequency, enhancing the decoupling effects. The latter stands for a propagating wave such $\omega < \alpha \rho_0$. Propagating magneto-acoustic waves exhibits a maximum dissipation for $\omega \approx \alpha \rho_0$ and it decreases for larger frequencies \citep[see discussions by][]{SolCarBal2013aa, PopLukKho2019ab}. For the waves we study, both effects are entangled, so it is not straightforward to talk about dissipation properly. Equations \eqref{eq:wequnz}, \eqref{eq:wequcz}, \eqref{eq:wequz} and \eqref{eq:wequambz} can be solved by variable separation. If the background variables are constant, then the wave variables admit a solution of the type $f(z,t) = \tilde{f}(z) e^{-i\omega t}$. The term proportional to the first spatial derivative is the one that drives the change of the wave amplitude along the atmosphere. To decouple this effect from the dissipation, we introduce the wave variables 
\citep[see similar derivations by][]{Lam1932aa, DeuGou1984aa},
\begin{gather}
    \psi_{\rm n,z} = (\rho_{\rm n,0} c_{\rm sn}^2)^{\frac{1}{2}} \tilde{u}_{\rm n,z} = a_{\rm n} \tilde{u}_{\rm n,z},\label{eq:psin}\\
    \psi_{\rm c,z} = (\rho_{\rm c,0}(c_{\rm sc}^2 + c_{\rm A,c}^2))^{\frac{1}{2}} \tilde{u}_{\rm c,z} = a_{\rm c} \tilde{u}_{\rm c,z},\label{eq:psic}\\
    \psi_{\rm z} = (\rho_{0} c_{\rm s}^2)^{\frac{1}{2}} \tilde{u}_{\rm z} = a_{\rm u} \tilde{u}_{\rm z},\label{eq:psi}\\
    \psi_{\rm Amb,z} = (\rho_0 c_{\rm A}^2)^\frac{1}{2} \tilde{u}_{\rm Amb,z} = a_{\rm Amb} \tilde{u}_{\rm Amb,z}.\label{eq:psiAmb}
\end{gather}

The transformations are introduced in Eqs. \eqref{eq:wequnz}, \eqref{eq:wequcz}, \eqref{eq:wequz} and, \eqref{eq:wequambz}, taking into account that the proportional factors are not a function of the time. Now, the $\psi$ functions only depend on $z$ and the time derivatives can be removed from the wave equations accordingly to the sinusoidal solution. Therefore, the equations for the 2F spatial variables $\psi$ read as (in 2F), 
\begin{subequations}
\begin{gather}
   \frac{d^2 \psi_{\rm n,z}}{dz^2} + \frac{\omega^2}{c_{\rm sn}^2}  \psi_{\rm n,z} = i \omega \alpha_0 \frac{\rho_{\rm c,0}}{c_{\rm sn}^2}\left(\frac{a_{\rm n}}{a_{\rm c}}\psi_{\rm c,z} - \psi_{\rm n,z}\right),\label{eq:weqpsin}\\
     \frac{d^2 \psi_{\rm c,z}}{dz^2} + \frac{\omega^2}{c_{\rm sc}^2 + c_{\rm Ac}^2}\psi_{\rm c,z} = - i \omega \alpha_0 \frac{\rho_{n,0}}{c_{\rm sc}^2 + c_{\rm Ac}^2}\left(\psi_{\rm c,z} - \frac{a_{\rm c}}{a_{\rm n}}\psi_{\rm n,z}\right),
     \label{eq:weqpsic}
\end{gather}
\end{subequations}
and those for the 1F model (see Eqs. \eqref{eq:wequz} and \eqref{eq:wequambz})
\begin{subequations}
\begin{gather}
    \frac{d^2 \psi_{\rm u,z}}{d z^2} + \frac{\omega^2}{c_{\rm s}^2} \psi_{\rm u,z} =\frac{i \omega}{\rho_0 c_{\rm s}^2\eta_{\rm A,0}}\left(\frac{a_{\rm u}}{a_{\rm Amb}} \psi_{\rm Amb,z} - \psi_{\rm u,z}\right),\label{eq:weqpsi}\\
    \frac{d^2 \psi_{\rm Amb,z}}{d z^2} = - \frac{i \omega}{c_{\rm A}^2 \rho_0 \eta_{\rm A,0}}\left(\psi_{\rm Amb,z} - \frac{a_{\rm Amb}}{a_{\rm u}}\psi_{\rm u,z}\right).
    \label{eq:weqpsiamb}
\end{gather}
\end{subequations}

We neglect the terms that involve the derivatives of $a_{\rm \alpha}$ terms, with $\alpha \in \{ \rm{n, c, u, Amb} \}$, consistently with the condition that the wavelength is much smaller than the pressure scale height.

We use the Eikonal approach to approximate the wave equation solution, assuming that the wave amplitude changes on a scale larger than its wavelength \citep[see e.g.,][]{Wei1962aa, DePHae1998aa, TraBriRic2014aa}. 
This means that a solution of the form $\vec{\psi} = A(z) e^{i \theta} \vec{e}(z)$ can be proposed, where $\vec{\psi}$ is the vector for the wave solutions (e.g., $\vec{\psi} = (\psi_{\rm n}, \psi_{\rm c})$ in the 2F model), $A(z)$ is the wave amplitude, $\theta = \int_{z_0}^z k_{z}(z') dz'$ is the spatial wave phase, with $k_{\rm z}$ the vertical wave number, and $\vec{e}$ is the polarisation vector. By neglecting spatial derivatives on $k_{\rm z}$, $A$ and $\vec{e}$,  \citep[see, e.g., ][]{DePHae1998aa}, the set of differential equations reduces to a set of polynomial equations of the form
\begin{equation}
    \hat{D}_{\rm 2F} \cdot \vec{e}_{\rm 2F} = 0;~~~ \hat{D}_{\rm 1F} \cdot \vec{e}_{\rm 1F} = 0,
\end{equation}
where, 
\begin{gather}
    \hat{D}_{2F} =
    \begin{pmatrix}
         k_{\rm z}^2 - \left(\frac{\omega^2}{c_{\rm sn}^2} +  i  \omega \alpha_0 \frac{\rho_{\rm c,0}}{c_{\rm sn}^2} \right) &  i \omega \alpha_0 \frac{\rho_{\rm c,0}}{c_{\rm sn}^2}\frac{a_{\rm n}}{a_{\rm c}}\\
       i \omega \alpha_0 \frac{\rho_{\rm n,0}}{c_{\rm sc}^2 + c_{\rm Ac}^2}\frac{a_{\rm c}}{a_{\rm n}} & k_{\rm z}^2 - \left(\frac{\omega^2}{c_{\rm sc}^2 + c_{\rm Ac}^2} +  i  \omega \alpha_0 \frac{\rho_{\rm n,0}}{c_{\rm sc}^2 + c_{\rm Ac}^2} \right)
    \end{pmatrix}, \label{eq:D2F}
    \\
    \hat{D}_{1F} =
    \begin{pmatrix}
        k_{\rm z}^2 - \left(\frac{\omega^2}{c_{\rm s}^2} + i \frac{\omega}{c_{\rm s}^2 \rho_0 \tilde{\eta}_{\rm A,0}}\right) &  i \frac{\omega}{c_{\rm s}^2 \rho_0 \eta_{\rm A,0}} \frac{a_{\rm u}}{a_{\rm Amb}}\\
         i \frac{\omega}{ c_{\rm A}^2 \rho_0 \tilde{\eta}_{\rm A,0}}\frac{a_{\rm Amb}}{a_{\rm u}} & k_{\rm z}^2 - i \frac{\omega}{c_{\rm A}^2 \rho_0 \eta_{\rm A,0}}
    \end{pmatrix}, \label{eq:D1F}
\end{gather}
are the dispersion matrices and, 
\begin{equation}
    \vec{e} = \frac{1}{\sqrt{\left|\frac{D_{12}}{D_{11}}\right|^2 + 1}} \left(- \frac{D_{12}}{D_{11}} , 1\right) \label{eq:polarisation}
\end{equation}
is the polarisation vector, which is defined as the unit null vector of the dispersion matrix. 

The determinant of the dispersion matrix must be equal to zero to obtain the non-trivial solution of the system of equations. This condition leads to the well-known dispersion relations (see \citep{KumRob2003aa, ZaqKhoRuc2011aa, SolCarBal2013ab, PopLukKho2019ab} for the 2F case and \citep{ChaCow1970aa, ForOliBal2007aa, BalAleCol2018aa} for the 1F one), which are in each case,
\begin{gather}
    \det(\hat{D}_{\rm 2F}) = k_{\rm z}^4 - \left[\frac{\omega^2}{c_{\rm sn}^2} + \frac{\omega^2}{c_{\rm sc}^2 + c_{\rm Ac}^2} + i  \omega \alpha_0 \left(\frac{\rho_{\rm c,0}}{c_{\rm sn}^2} + \frac{\rho_{\rm n,0}}{c_{\rm sc}^2 + c_{\rm Ac}^2}\right)\right]k_{\rm z}^2 \nonumber \\
    + \frac{\omega^4}{c_{\rm sn}^2(c_{\rm sc}^2 + c_{\rm Ac}^2)} + i \omega^3 \alpha_0 \left(\frac{\rho_{\rm n,0} + \rho_{\rm c,0}}{c_{\rm sn}^2(c_{\rm sc}^2 + c_{\rm Ac}^2)}\right) = 0,\label{eq:detD2F}\\
    \det(\hat{D}_{\rm 1F}) = k_{\rm z}^4 - \left[\frac{\omega^2}{c_{\rm s}^2} + i \frac{\omega}{\rho_0 \tilde{\eta}_{\rm A,0}}\left(\frac{1}{c_{\rm s}^2} + \frac{1}{c_{\rm A}^2}\right)\right]k_{\rm z}^2 + i \frac{\omega^3}{c_{\rm s}^2 c_{\rm A}^2 \rho_0 \tilde{\eta}_{\rm A,0}} = 0. \label{eq:detD1F}
\end{gather}

Both equations are 4th order in wavenumber ($k_{\rm z}$), leading to two different pairs of normal modes, each pair containing an upward and a downward propagating mode. We observe that $\det(\hat{D}_{\rm 2F})$ contains a term proportional to $\omega^4$ that is missing in the $\det(\hat{D}_{\rm{1F}})$ expression. From Eq. \eqref{eq:detD2F}, the ratio of the $\omega^4$-proportional term to the $\omega^3$-proportional one, in absolute value, is $\omega / (\rho_0 \alpha_0)$.
This ratio is nothing but the estimator for the contribution of the inertial terms in the drift velocity equation (see Eq. \eqref{eq:Est_inertia}).
Therefore, the absence of the $\omega^4$ term is the consequence of neglecting inertial terms in Eq. \eqref{eq:w_1F}, as noted by \citet{ZaqKhoRuc2011aa} for the case of Alfvén waves. 

\begin{figure}[!t]
\centering
\includegraphics[width=\hsize]{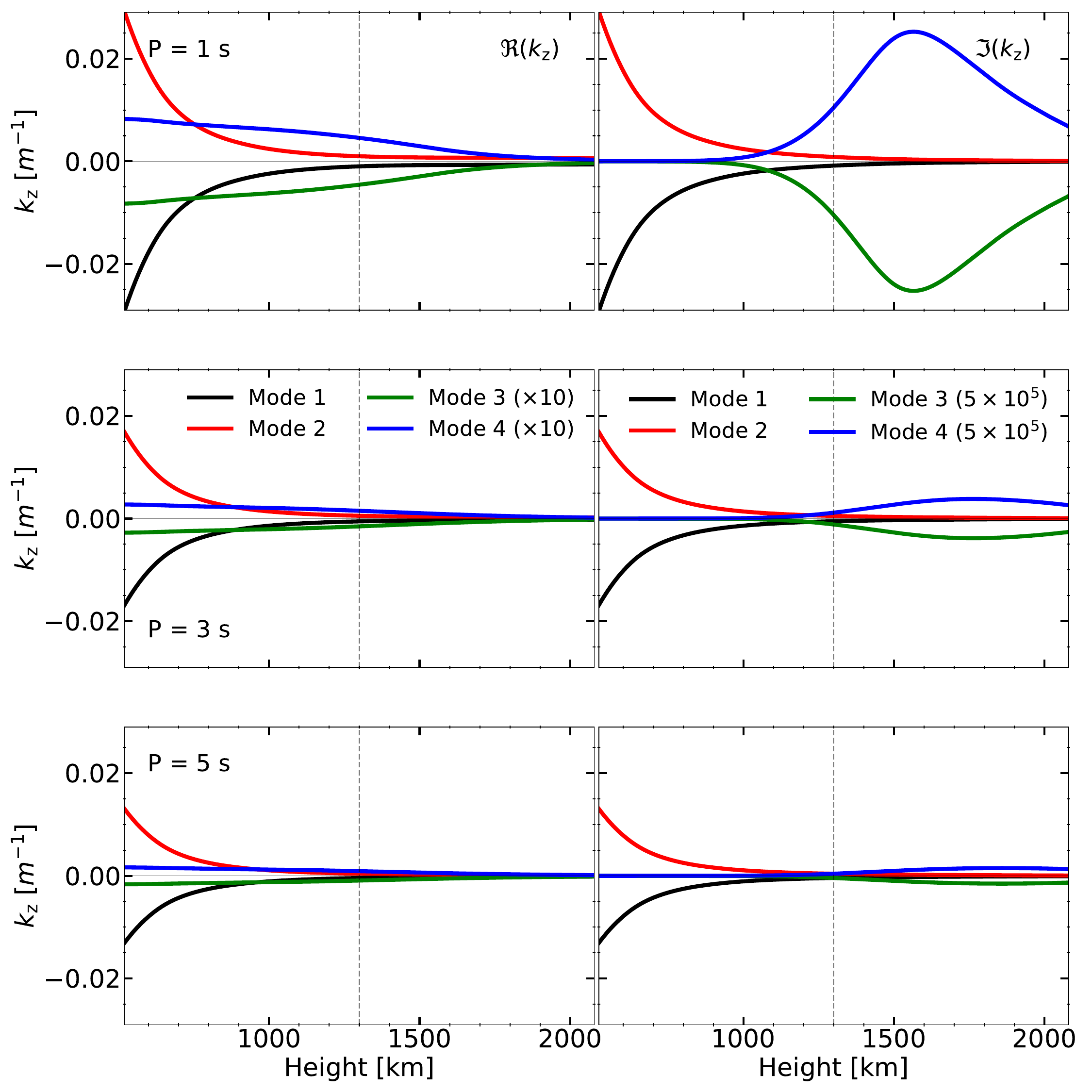}
\caption{Vertical wavenumbers for the 2F model as functions of height for all wave-periods studied in this work. From the top to the bottom, the wave-periods are 1 s, 3 s and 5 s. Blue and green solid lines: upward and downward fast mode; red and black solid lines: upward and downward normal mode (different from the fast mode). The left column represents the real part, and the right represents the imaginary part.   The fast mode lines are multiplied by an arbitrary factor for the representation. The vertical dashed line indicates the equipartition layer $\beta_\text{plasma} = 1$.}
\label{fig:disprel}
\end{figure}

Working in terms of the wave functions described by Eqs. \eqref{eq:psin}, \eqref{eq:psic}, \eqref{eq:psi}, and \eqref{eq:psiAmb}, rather than the model variables $(u_{\rm n}, u_{\rm c}, u, B_{\rm x})$ offers the advantage of symmetry between upward and downward modes, and the imaginary part of the wave number is directly linked to the collisional dissipation. In Fig. \ref{fig:disprel}, we show the solutions of the dispersion relations given by Eqs. \eqref{eq:detD2F} and \eqref{eq:detD1F} for our model atmosphere.
 
One pair of solutions has a similar behaviour as the classic fast magneto-acoustic mode, but it dissipates energy because of charge-neutral collisions (mode 4 travels upwards and mode 3, downwards). As we have seen before, at deep layers the fast mode behaves as an acoustic wave, travelling at the speed $c_{\rm s} \sim c_{\rm sn}$, since the atmosphere is weakly ionized. Once the wave crosses the equipartition layer, it behaves as a magnetic wave, travelling with a speed that tends to $c_{\rm A}$ when reaching layers where $\beta_{\rm plasma} \ll 1$. The other mode (mode 2 traveling upwards and mode 1, downwards) is highly dissipated at the deep layers as the real and imaginary parts of $k_{\rm z}$ have similar values. These solutions resemble those presented in Fig. 13 of \citet{PopLukKho2019ab}, but here we have included all the wave-periods explored in this work.

\begin{figure}[!t]
\centering
\includegraphics[width=\hsize]{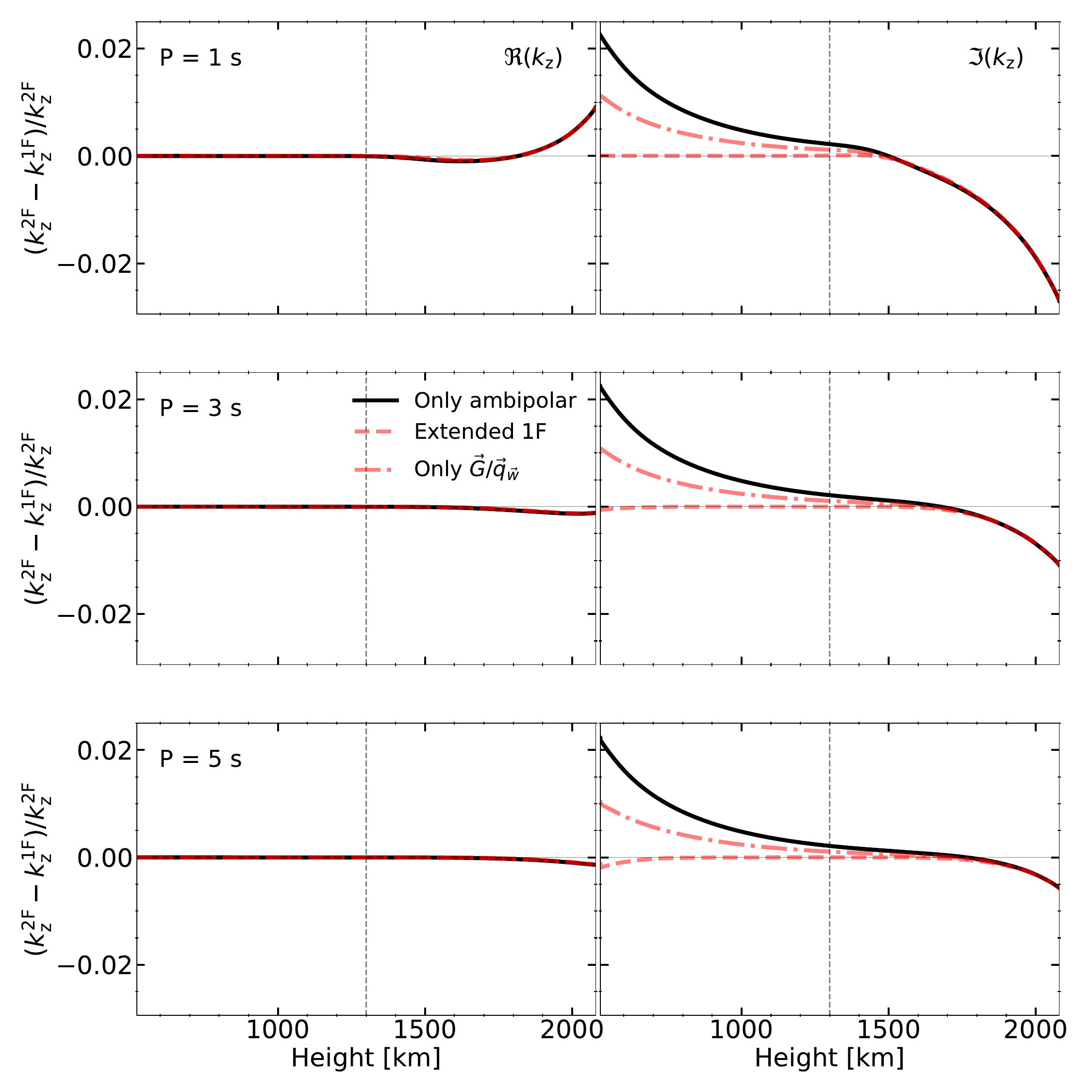}
\caption{Difference between the 2F and the 1F wavenumbers relative to the 2F wavenumber. Left column: calculations for the real part of $k_{\rm z}$; right column: same calculations, but for the imaginary part of $k_{\rm z}$. Only the upward fast mode solution is represented. Solid black line: differences computed asumming the 1F model described in section \ref{sec:1F}; dashed red line: differences computed extending the 1F for accounting the contributions of $\vec{q}_{\rm w}$ and $\vec{G}$; dashed-dotted red line: differences computed extending the 1F for accounting the contributions of $\vec{q}_{\rm w}$ alone and $\vec{G}$ alone. The vertical dashed line indicates the equipartition layer $\beta_\text{plasma} = 1$.}
\label{fig:reldiff}
\end{figure}

For model comparison, we show in Fig. \ref{fig:reldiff} (black solid line) the relative difference between the 1F and 2F wavenumbers with respect to the value computed for the 2F model, for both the real and imaginary parts. At this point of the discussion, we focus on the black curve, letting the others for section \ref{sec:qsmall}. We only excite the fast mode in the numerical experiments, so we focus the discussion on this mode. Only the upwards mode is represented. Differences in the real part of $k_{\rm z}$ are negligible at deep layers (of the order of 0.1\%), reaching a maximum of 1\% at the high layers for the 1 s wave-period. As expected, those differences increase with the wave frequency. However, the imaginary part of $k_{\rm z}$ shows more significant differences at deep and high layers. We note that the differences at the high layers increase with the wave frequency, but those at deep layers seem to be unaffected by the wave period. This result shows that the waves are more damped in the 2F model at deep layers, but the opposite occurs at high layers. This picture matches qualitatively with the temperature increase residuals shown in Fig. \ref{fig:heating}. Even though Fig. \ref{fig:wvsw} informs us that the non-linear effects play an important role, Fig. \ref{fig:reldiff} gives us a chance to figure out an explanation for model discrepancies using the linear theory.

\begin{figure}[!t]
\centering
\includegraphics[width=\hsize]{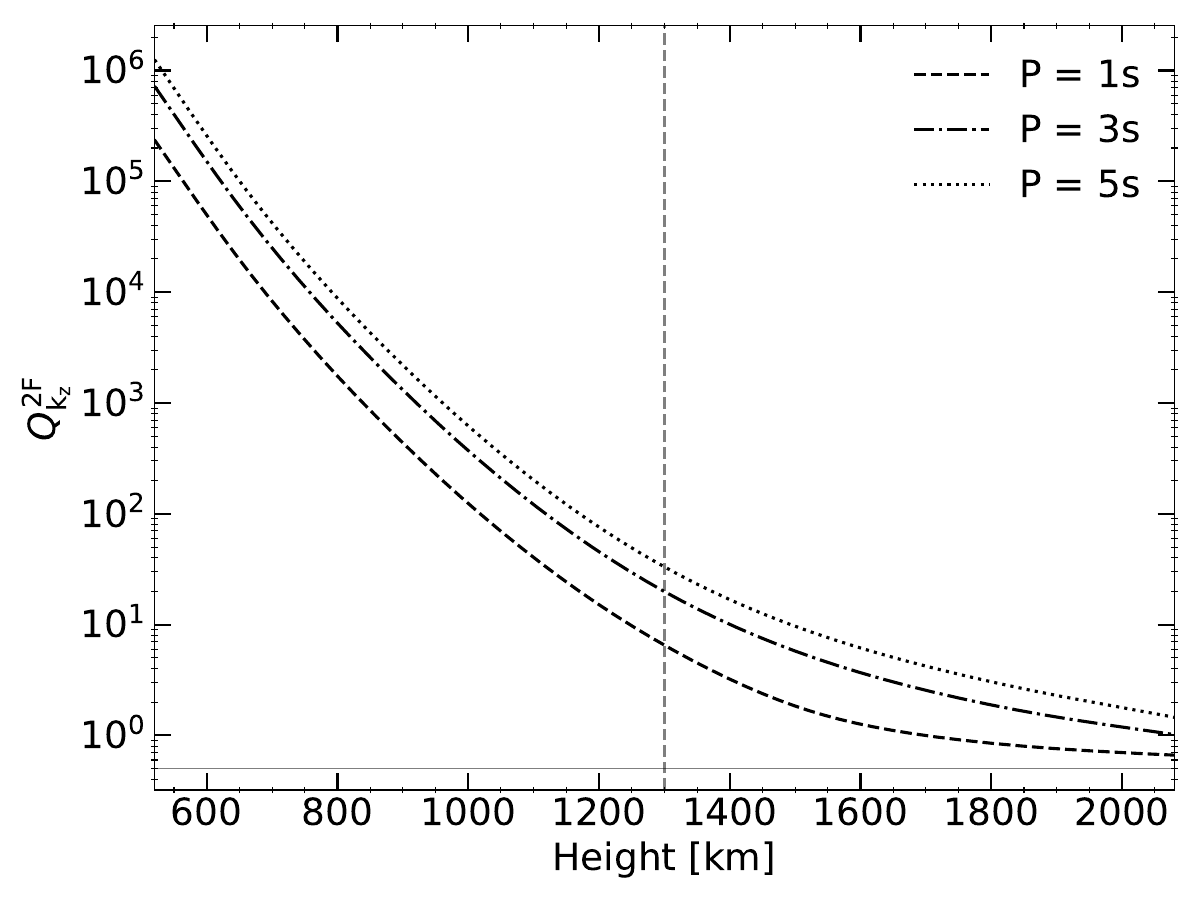}
\caption{Quality factors for the simulated wave-periods for the 2F model. The horizontal grey line indicates the value $Q^{\rm 2F}_{k_{\rm z}} = \frac{1}{2}$. The vertical dashed line indicates the equipartition layer $\beta_\text{plasma} = 1$.}
\label{fig:Quality}
\end{figure}

A useful parameter to put into context the importance of the differences in $k_{\rm{z}}$ that we have obtained is the quality factor of the waves, commonly defined as
\begin{equation}
Q_{k_{\rm z}} = \frac{1}{2}\left|\frac{\Re(k_{\rm z})}{\Im(k_{\rm z})}\right|. \label{eq:Quality}
\end{equation}
If $Q_{k_{\rm z}} = \frac{1}{2}$, the wave is effectively damped in a distance of a wavelength. For $Q_{k_{\rm z}} > \frac{1}{2}$, the wave dissipates energy in scales larger than its wavelength and for $Q_{k_{\rm z}} < \frac{1}{2}$, it is efficiently dissipated before reaching the scale of its wavelength. Fig. \ref{fig:Quality} shows that below 1000 km, dissipating the wave energy requires more than 100 periods, indicating a weak dissipation. However, non-linear effects become important for the 1 s and 3 s period waves at that height, enhancing energy dissipation, as shown in Figs. \ref{fig:wvsw}, \ref{fig:energy1F} and \ref{fig:energy2F}. As the waves reach layers of $\beta_{\rm plasma} < 1$, the wavelength is lengthened and the imaginary part of $k_{\rm z}$ increases, so we get $Q \approx 1$ at high layers (see Fig. \ref{fig:disprel}). Fig. \ref{fig:reldiff} indicates that the more significant differences appear at the bottom of the atmosphere, where the waves propagate almost without dissipation or at the top layers, where the waves arrive with low energy. Therefore, the differences in wave flux shown in Fig. \ref{fig:ediff} may be caused by deviations at the intermediate layers.


\section{On the model discrepancies \label{sec:discrep}}

In the previous sections, we have compared a purely hydrogen 2F model with a standard 1F model with ambipolar diffusion, whose parameters $\xi_{\rm n},~ \tilde{\mu},~ \tilde{\eta}_{\rm A}$ are introduced consistently for such plasma. After examining numerical simulations and analytical solutions, we find deviations in the imaginary part of $k_{\rm z}$ between models to explain why the wave in the 1F model is transporting upwards more energy than in the 2F model. Now, we try to determine the origin of those differences by looking into the model construction. As the main assumptions of the 1F model are listed in Section \ref{sec:1F}, here we delve into the variable definitions.

While the frame of reference in NF models refers, say, to the laboratory frame of reference, the 1F variables refer to the centre of mass frame of reference, so their definitions depend on the relative velocities of each component to the centre of mass velocity, $\vec{w}_\alpha$.
In Appendix \ref{App:B}, we list the definitions of the 1F variables in terms of the 2F variables. For further details, we refer the reader to the works by \citet{Bra1965aa}, \citet{Sch1977aa} or \citet{KhoColDia2014aa}, among others. The inspection of the definitions bring out that most of them involve terms proportional to $\rho_\alpha w_{\alpha}^2$ or $\rho_\alpha w_{\alpha}^3$, which can be neglected in our study (see discussion above Eq. \eqref{eq:T_1F}). However, the 1F heat flux (see Eq. \eqref{eq:q_t}) includes two terms proportional to $\vec{w}_\alpha$ and to the thermal pressure. The latter was also noted by \citet{Sch1977aa}, who suggested checking these terms when comparing models with different variable definitions in the velocity frame of reference.  The 1F heat flux components proportional to $\vec{w}_\alpha$ only appear when it has been derived from a NF model closed without including a heat flux for each species, as the 2F model discussed here. In such situation, as shown in Appendix \ref{App:Ext1F}, the terms regarding the 1F heat flux definition appear, even though there is not any heat flux present in the NF model. From Eqs. \eqref{eq:q_t}, \eqref{eq:wn} and \eqref{eq:wc}, we find this 1F heat flux to be 
\begin{equation}
     \vec{q}_{\rm w} \coloneqq \frac{\gamma}{\gamma - 1} \sum_{\{\rm n,c\}} p_\alpha \vec{w}_\alpha \approx \frac{\gamma}{\gamma - 1} \xi_{\rm n} (1 - \xi_{\rm n})\,\tilde{\mu}\, p\, \vec{w}, \label{eq:q1F}
\end{equation}
in absence of viscosity. We point to the second expression when referring to $\vec{q}_{\rm w}$, which is derived from the previous one under the assumption of  $T_{\rm n} = T_{\rm c}$, as in the case of $\vec{G}$ (see Eq. \ref{eq:G_novisc}). In addition, as for the $\vec{G}$ term, it vanishes for the high and low ionisation limits. Both $\vec{S}^*$ (or $Q$) and $\vec{q}_w$ are of the order $\mathcal{O}(w)$, so the last one seems a valuable candidate to take into account for explaining the model discrepancies.  

\subsection{\label{sec:qsmall} A look into the neglected contributions}
 
We have now discussed the main approximations considered in the 1F modelling. Here, we examine the ones we consider that mainly influence the discrepancies between models: the contributions of $\vec{G}$, $\vec{q}_{\rm w}$ and the inertial terms. For the first two cases, Eqs. \eqref{eq:G_novisc} and \eqref{eq:q1F} allow us to incorporate these effects in the 1F model described in section \ref{sec:1F}. For the case of the inertial terms, it is not straightforward to account for them easily in the 1F model, but we can use Eq. \eqref{eq:Est_inertia} for estimating them.

We consider first the contributions of $\vec{G}$ and $\vec{q}_{\rm w}$ to the dispersion relation. For the sake of simplicity, we directly neglect the terms involving the derivatives of the background quantities, which is reasonable for the range of frequencies of this study. The mass and momentum equations remain as before. We leave the details about how to extend the energy and induction equation for Appendix \ref{App:Ext1F}. From the set of Eqs. \eqref{eq:rho_lin_ext}, \eqref{eq:mom_lin_ext}, \eqref{eq:e_lin_ext} and \eqref{eq:induc1F_lin_ext}, we obtain the dispersion relation
\begin{gather}
    k_{\rm z}^4 - (1 + \xi_{\rm c,0})^2\left[\frac{\omega^2}{c_{\rm s}^2} + \frac{\omega^2 \xi_{\rm c,0}^2 \tilde{\mu}_0^2}{c_{\rm A}^2}+ i \frac{\omega}{\rho_0 \tilde{\eta}_{\rm A,0}}\left(\frac{1}{c_{\rm s}^2} + \frac{1}{c_{\rm A}^2}\right)\right]k_{\rm z}^2 \nonumber \\ + i(1 + \xi_{\rm c,0})^2\frac{\omega^3}{c_{\rm s}^2 c_{\rm A}^2 \rho_0 \tilde{\eta}_{\rm A,0}} = 0.
    \label{eq:dispqw}
\end{gather}

As expected, $\vec{G}$ and $\vec{q}_{\rm w}$ slightly modify the expression derived in Eq. \eqref{eq:detD1F}. We solve Eq. \eqref{eq:dispqw} for the upward fast mode and represent it in Fig. \ref{fig:reldiff} in red dashed lines. We refer to this model as the extended 1F model. We find that the new dispersion relation can solve the discrepancies between models at deep and intermediate layers. As we have noted before (see discussion in section \ref{sec:eikonal} about Fig. \ref{fig:reldiff}), those differences look weakly dependent on the wave period. For the model atmosphere, we have $\frac{\xi_{\rm c,0}^2}{c_{\rm A}^2} = \frac{1}{c_{\rm A,c}^2} \ll \frac{1}{ c_{\rm s}^2}$ and the terms proportional to $\omega^2$ in the coefficient of $k_{\rm z}^2$ in Eq. \eqref{eq:dispqw} reduce to the ones in Eq. \eqref{eq:detD1F}. Therefore, the modifications introduced by $\vec{G}$ and $\vec{q}_{\rm w}$ (proportional to $1 + \xi_{\rm c,0}$) are appropriate for explaining those differences. As closure, if one only considers the term $\vec{G}$ or the term $\vec{q}_{\rm w}$ (only the ambipolar contribution) alone, it follows
\begin{gather}
    k_{\rm z}^4 - (1 + \xi_{\rm c,0})\left[\frac{\omega^2}{c_{\rm s}^2} + i \frac{\omega}{\rho_0 \tilde{\eta}_{\rm A,0}}\left(\frac{1}{c_{\rm s}^2} + \frac{1}{c_{\rm A}^2}\right)\right]k_{\rm z}^2 \nonumber \\ + i(1 + \xi_{\rm c,0})\frac{\omega^3}{c_{\rm s}^2 c_{\rm A}^2 \rho_0 \tilde{\eta}_{\rm A,0}} = 0.
    \label{eq:dispqw2}
\end{gather}
This dispersion relation is a particular case of the one derived by \citet{ForOliBal2007aa}, when setting $k_{\rm x} = 0$ and in absence of collisions with electrons. We also represent this case in Fig. \ref{fig:reldiff}, finding that both terms are needed for the correct match between the imaginary part of the wavenumbers.

\begin{figure}[!t]
\centering
\includegraphics[width=\hsize]{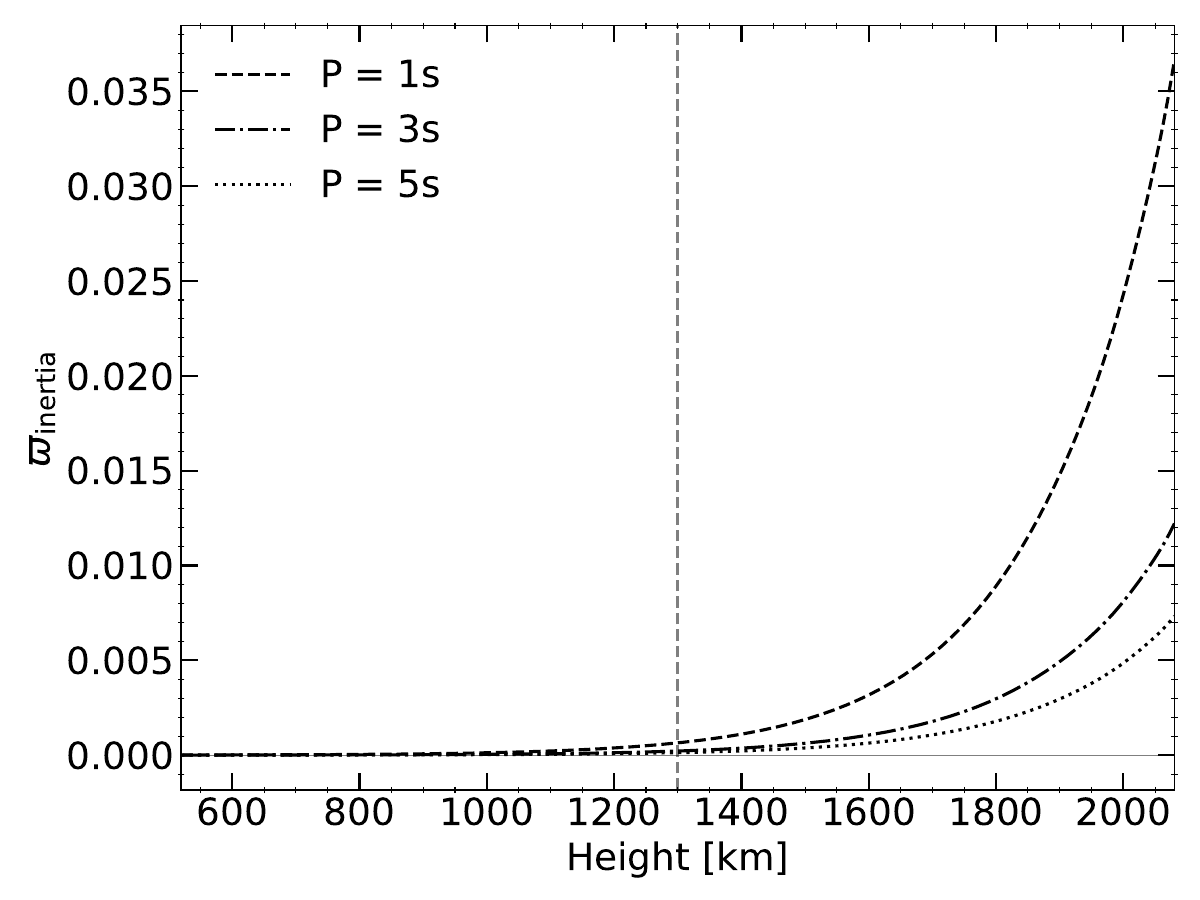}
\caption{Estimator of the impact of inertial terms for the three wave-periods considered in this work as function of the height. The vertical dashed line indicates the equipartition layer $\beta_\text{plasma} = 1$.}
\label{fig:estimators}
\end{figure}

After adding the corrections of our extended 1F model, the period dependent discrepancies at high layers remain. In Fig. \ref{fig:estimators} we represent the estimator for the inertial terms for all wave periods computed after Eq. \eqref{eq:Est_inertia}. We find them to be negligible at the deep layers, where the collision frequency is large, but they become more relevant at the high layers. By comparing $\varpi_{\rm inertia}$ with the curves for the extended 1F model in Fig. \ref{fig:reldiff}, we find the same trend with height and period, and the values are comparable. As mentioned, we cannot provide a direct proof confirming that inertial terms produce those differences at high layers, however, Fig. \ref{fig:reldiff} tentatively points to such conclusion.

It must be noted that the above discussion is limited to the model atmosphere considered in the present work, which has a rather low ionisation fraction. For high ionisation fractions, it can be expected that both magnetic and thermal pressure decoupling mechanisms vanish  \citep[see Eq. \eqref{eq:fullw} or Eq. 29 in][]{ForOliBal2007aa} and the heating is significantly reduced. Therefore, the discrepancies between 2F and 1F models would be also reduced in absolute terms. In the intermediate situation, when neutrals and charges have similar numbers,  both decoupling mechanisms become of the same order of magnitude \citep[see discussion by][]{PinGalBac2008aa}. The drift velocities are expected to be smaller than in the weakly ionised atmosphere, but, nevertheless, non negligible.

\subsection{Implications on the non-linear regime \label{sec:non-linear}}

We have presented above an argument for solving the discrepancies between models in the linear regime. Now, we take a look back to the simulations and the non-linear regime. 

We focus again on the internal energy equations. We have established that the main discrepancies between models are driven by the intermediate layers, so we consider the approximation Eq. \eqref{eq:w_1F} to the drift velocity and leave the inertial terms aside. Then, we take Eq. \eqref{eq:QnQc} for the 2F model and $Q = \vec{J} \cdot \vec{E}^*$ for the 1F model, so,
\begin{subequations}
\begin{gather}
    Q_{\rm n} + Q_{\rm c} = \frac{\tilde{\eta}_{\rm A}}{\xi_{\rm n}^2} [\xi_{\rm n}^2(\nabla \cdot \hat{p}_{\rm m})^2 + 2 \xi_{\rm n} \vec{G} \cdot (\nabla \cdot \hat{p}_{\rm m}) + G^2], \label{eq:QnQcw1F}\\
    Q = \frac{\tilde{\eta}_{\rm A}}{\xi_{\rm n}^2} [\xi_{\rm n}^2(\nabla \cdot \hat{p}_{\rm m})^2 + \xi_{\rm n} \vec{G} \cdot (\nabla \cdot \hat{p}_{\rm m})]. \label{eq:Qw1F}
\end{gather}
\end{subequations}

We have that Eq. \eqref{eq:QnQcw1F} coincides with the expression obtained by \citet{Bra1965aa}, \citet{KhoArbRuc2004aa} or \citet{KhoRucOli2006aa}, and the first order expansion for high collisional coupling derived by \citet{BenFli2020aa}.
We obtain that $Q_{\rm n} + Q_{\rm c} \neq Q$. Nevertheless, the equality is fulfilled if $\vec{G}$ is neglected. On one side, Eq. \eqref{eq:QnQcw1F} indicates that including the $\vec{G}$ term increases the heating. On the other side, Eq. \eqref{eq:Qw1F} is not strictly positive if $\vec{G} \cdot (\nabla \cdot \hat{p}_{\rm m}) < 0$ and $G > |\nabla \cdot \hat{p}_{\rm m}|$. The first condition may depend on the phenomena (waves, convection, Rayleigh-Taylor instability, ...) and the second condition stands when the $\vec{G}$ term dominates \citep[see discussions by][]{Bra1965aa, KhoArbRuc2004aa, KhoRucOli2006aa, PinGalBac2008aa}. As mentioned in Sec. \ref{sec:1F}, the $\vec{G}$ term is not of interest in many practical situations, so $Q$ is given by Eq. \eqref{eq:Qamb} in a good approximation, but \eqref{eq:Qw1F} might be smaller than the former if the first condition is fulfilled. For instance, we have from Eqs. \eqref{eq:q_1F_2} and \eqref{eq:qw_lin} that $\vec{G} \cdot (\nabla \cdot \hat{p}_{\rm m}) > 0$ in the linear regime, so we expect $Q$ to be larger when $\vec{G}$ is included. Therefore, the 1F heating at intermediate layers can be enhanced for our setup if $\vec{G}$ is included.

Finally, we also exploit the advantages of accounting for $\vec{q}_{\rm w}$. This term appears as a flux term in the energy equation so, in principle, it plays no role in the local heating. In Eq. \eqref{eq:q1F}, we have put together the contributions from the internal energy flux and the pressure flux. Now, let us split them in $\vec{q}_{\rm w}^{(1)} = \frac{1}{\gamma - 1} \sum p_\alpha \vec{w}_\alpha$ and $\vec{q}_{\rm w}^{(2)} = \sum p_\alpha \vec{w}_\alpha$. Then, from Eqs. \eqref{eq:eq_e_ext} and \eqref{eq:mom_ext}, we have
\begin{equation}
    \frac{\partial e_{\rm int}}{\partial t} + \nabla \cdot (e_{\rm int} \vec{u} + \vec{q}_{\rm}^{(1)}) = - \nabla \cdot \vec{q}_{\rm}^{(2)} - p \nabla \cdot \vec{u} + Q. \label{eq:eint_ext}
\end{equation}
Next, we split $\nabla \cdot \vec{q}_{\rm}^{(2)} = \sum p_\alpha \nabla \cdot \vec{w}_\alpha + \sum \vec{w}_\alpha \cdot \nabla p_\alpha$. From the second term, considering the relations \eqref{eq:wn}, \eqref{eq:wc}, and \eqref{eq:G}, it follows
\begin{equation}
    \sum_{\{n,c\}} \vec{w}_\alpha \cdot \nabla p_\alpha = - \vec{w} \cdot \vec{G} \label{eq:QQ1F_1}
\end{equation}
If one sums Eqs. \eqref{eq:Qw1F}, which reads as $\xi_{\rm n}\vec{w} \cdot (\nabla \cdot \hat{p}_{\rm m})$, and \eqref{eq:QQ1F_1}, then Eq. \eqref{eq:QnQcw1F} is recovered. This result aligns with \citet{Bra1965aa} and demonstrates that including $\vec{G}$ together with $\vec{q}_{\rm w}$ is required for recovering the exact 2F heating rate, under the assumption of neglecting the inertial terms. Once again, this conclusion agrees with the discussion in section \ref{sec:qsmall} in the linear regime, where we find that adding $\vec{G}$ or $\vec{q}_{\rm w}$ alone is not sufficient for solving model discrepancies. In addition, when including $\vec{q}_{\rm w}$, we get extra terms: $\nabla \cdot \vec{q}^{(1)}$ is related with the internal energy flux and  $\sum p_\alpha \nabla \cdot \vec{w}_\alpha$ with the adiabatic source term (but this term is not adiabatic). The first term can be easily adapted for the 1F model following Eq. \eqref{eq:q1F}, but the second one leads to complicate expressions. Therefore, we would suggest to apply a 2F modelling for a practical implementation of those effects, if it is required.

\section{Conclusions \label{sec:Conclusions}}

In this work, we compare two models that apply for a partially ionized atmosphere: a 2F model composed by neutrals and charges \citep{Bra1965aa, ZaqKhoRuc2011aa, MeiShu2012aa, SolCarBal2013aa, SolCarBal2013ab, LeaLukLin2013aa, KhoColDia2014aa, PopLukKho2019aa, PopLukKho2019ab}, and a 1F model that only includes ambipolar diffusion as a non-ideal effect \citep{ForOliBal2007aa, KhoCol2012aa, KhoCol2012ab, SolCarBal2015aa,CalKho2019aa, KhoCal2019aa, PopKep2021aa} The extensively used 1F model describes  the interaction between charges and neutrals through the ambipolar diffusion term in the generalized Ohm's law. This term depends on the drift velocity between charges and neutrals. To analyse the heating process, we obtain the heating source terms for each model. 
Our research focuses on estimating the heating caused by wave propagation in the atmosphere using specific models. Similar to the study by \citet{PopLukKho2019ab}, we establish a hydrostatic equilibrium atmosphere with a VAL3C temperature profile and a uniform horizontal magnetic field. The strength of the magnetic field is chosen such that the equipartition region, $\beta_{\rm plasma}=1$, is located in the middle of the atmosphere. We simulate the propagation of small amplitude waves in this equilibrium atmosphere, with periods 1 ,3 and 5 seconds, from the bottom to the top, with a decreasing collision frequency along the wave path. From the simulations, we calculate the temperature increase due to the collisional heating. Additionally, we compare the 1F and 2F heating terms and analyse the distribution of the wave energy in each case. We conclude the analysis by employing the Eikonal approach to solve the wave equation. This allows us to derive the 1F and 2F dispersion relations for each simulated wave and then we compare the wavenumbers obtained for each model.

Below, we summarize the main findings of the research:
\begin{itemize}
    \item Similar to what was concluded before by many authors \citep[see, e.g.,][]{KumRob2003aa, ZaqKhoRuc2011aa, ZaqKhoSol2013aa, SolCarBal2013aa, SolCarBal2013ab, BalAleCol2018aa, PopLukKho2019ab, ZhaPoeLan2021aa, Sol2024aa}, MHD wave dissipation by means of elastic collisions is enhanced for short-period waves. However, as suggested by \citet{Ost1961aa} and in line with the works by, namely, \citet{PopLukKho2019ab} or \citet{ZhaPoeLan2021aa}, we find long-period waves to increase more significatively the background temperature. Waves less affected by collisional dissipation can carry and deposit more energy at the high layers of low collisional coupling. While the ratio of wave frequency to collision frequency is important, it is also essential to consider how effectively waves transport energy along their path.
    \item The differences between 
    the 2F and 1F temperature increase that we obtain align with those for the heating terms and the total energy flux of the waves. The latter reveals that the waves in the 2F model transports less energy upwards than in the 1F model. Since the 1F model allows the wave for transporting more energy to high layers, where the collision frequency is smaller, it can rise more the local temperature, which explains the results in Fig. \ref{fig:heating}. To check this hypothesis, we apply the linear theory for obtaining the dispersion relation for each model. The comparison of the wavenumbers for the fast mode brings out that the 2F model dissipates more energy than the 1F at deep and intermediate layers, but the opposite at the high layers. Therefore, the conclusions of the linear solutions align with those of the numerical experiments.
    \item We have explored the main assumptions that make the models different, finding the contribution of $\vec{G}$, and $\vec{q}_{\rm w}$ (the latter, related with the change in the reference system of velocities from that of neutrals and charges to the common centre of mass) as the cause of the discrepancies at deep layers. We also find the inertial terms to explain the discrepancies at the high layers.

    \item Finally, we demonstrate that $\vec{G}$ and $\vec{q}_{\rm w}$ are required for recovering the 2F heating rates in agreement with the findings of \citet{Bra1965aa}.
\end{itemize}

This study employed several assumptions to simplify the comparison between models. They proved helpful in delving into the physical details of the 1F approximations.

One of the primary limitations concerns the ionisation fraction, as neglecting ionisation and recombination effects means that each fluid balances itself alone in the equilibrium. Consequently, the variation in the ionisation fraction is solely determined by the different stratification of each fluid. \citet{ZhaPoeLan2021aa} highlighted the potential significance of the ionisation fraction in accounting for frictional heating, given the sensitivity of collision frequencies to this factor. When conducting 1F radiative MHD simulations, \citet{NobMorMar2020aa} found that the local thermodynamical (LTE) approximation significantly underestimates the ionisation fraction by 2-3 orders of magnitude in non-equilibrium (NEQ) flux emergence simulations, leading to an overestimation of ambipolar heating. In line with the previous work, \citet{MarCruGos2023aa} also found NEQ effects to be important at the upper layers of the chromosphere, allowing for a more significant temperature growth due to ambipolar heating. In the two fluid framework, \citet{SnoDruHil2023aa} studied shock propagation in the solar chromosphere accounting for ionisation and recombination processes, finding NEQ effects to reduce the temperature increase in the post-shock region when compared with MHD models.

Another improvement of the study would be including helium into the discussion. Some works have explored NF models that include helium  \citep[see, e.g.][]{ZaqKhoRuc2011ab, VraKrs2013aa, MarSolTer2017aa}. These works found the coupling between helium and hydrogen to be weaker than the coupling between neutral hydrogen and protons for high chromosphere and prominence conditions. As a consequence, they found larger drift velocities and damping effects if only hydrogen were considered, for the same wave frequencies. In this sense, it can be said that the helium reduces the applicability of 1F models.

Our work focuses on the ion-neutral collision damping mechanism (in 2F) and on the ambipolar diffusion mechanism (in 1F), and on their comparison. Therefore, we purposely reduced the amplitudes of the initial perturbation to prevent shock formation. In a more realistic situation, other damping mechanisms would be at play, such as viscosity, or Ohmic resistivity. These mechanisms might overcome the ambipolar mechanism \citep[see, eg.][]{TerMolWie2002aa, KhoArbRuc2004aa, CarTerOli2006aa, KhoRucOli2006aa, ForOliBal2008aa, SolCarBal2015aa}. The presence of non-linearities would enhance the energy dissipation due to ion-neutral collisions at shock wave fronts \citep{SnoHil2021aa, SnoDruHil2023aa}. Moreover, adiabatic heating in shocks can become more important than heating due to dissipation of shocks by ion-neutral collisions \citep{ArbBraShe2016aa, ZhaPoeLan2021aa}. Future studies would need to perform a thorough comparison of different dissipation mechanisms in the 1F and 2F models. Nevertheless, it must be taken into account that fast waves propagate at approximately the Alfv\'en speed in the $\beta_{\rm plasma} < 1$ and they are less likely to develop shock waves, which makes ion-neutral collision dissipation mechanisms important even for waves with larger amplitudes.

\begin{acknowledgement}
This work was supported by the European Research Council through the Consolidator Grant ERC-2017-CoG-771310-PI2FA and by Spanish Ministry of Science through the grant PID2021-127487NB-I00. M.M.G.M. acknowledges support from the Spanish Ministry of Science and Innovation through the grant CEX2019-0000920-S-20-1 of the Severo Ochoa Program. We thankfully acknowledge the resources of LaPalma Supercomputer, located at the Instituto de Astrofisica de Canarias.
\end{acknowledgement}


\bibliographystyle{bibtex/aa} 
\bibliography{MMGM} 

\begin{appendix}

\section{\label{App:A}Driver configuration}
We provide a brief overview of the setup for the driver located at the base of the atmosphere. This setup closely resembles the one utilized by \citet{PopLukKho2019aa}. Under photospheric conditions, a 1F solution effectively characterizes the waves in that region. We anticipate a strong coupling, indicating that $\vec{w}$ tends towards 0, and neutrals and charges can be assumed to move at the same velocity. 

The most straightforward approach is to follow the 1F simulation driver. From the linear theory, ideal MHD solution $\left\{\frac{u_{\rm z}}{c_{\rm s}}, \frac{\rho_1}{\rho_0}, \frac{p_1}{p_0}, \frac{B_{\rm x,1}}{B_{\rm x,0}}\right\} = \{U_{\rm z}, R, P, \tilde{B}_{\rm x}\} e^{i(k_{\rm z} z - \omega t)}$ is shown to fulfil the following relations 
\begin{gather}
    U = M, \label{eq:U}\\
    R = \frac{U c_{\rm s}}{\omega}\left( k_{\rm z} + \frac{i}{H}\right), \label{eq:R}\\
    P = \frac{\gamma U c_{\rm s}}{\omega} \left(k_{\rm z} + \frac{i}{\gamma H}\right), \label{eq:P}\\
    \tilde{B}_{\rm x} = \frac{U c_{\rm s}k_{\rm z}}{\omega}, \label{eq:Bx}
\end{gather}
where we choose the Mach number $M = 10^{-3}$ for all the simulations.

For the 2F simulation, the driver variables $\left\{\frac{u_{\rm \alpha,z}}{c_{\rm s}}, \frac{\rho_{\alpha,1}}{\rho_{0}}, \frac{p_{\alpha,1}}{p_{0}}, \frac{B_{\rm x,1}}{B_{\rm x,0}}\right \} = \{U_{\rm{\alpha}}, R_\alpha, P_\alpha, \tilde{B}_{\rm x}\}e^{i(k_{\rm z} z - \omega t)}$ are related as follows \citep[for further details, see][]{PopLukKho2019ab}: 
\begin{gather}  
    U_{\rm n} = U ; U_{\rm c} = U, \label{eq:Ualpha}\\
    R_{\rm n} = \frac{\rho_{\rm n,0}}{\rho_0} R; ~~~R_{\rm c} = \frac{\rho_{\rm c,0}}{\rho_0} R, \label{eq:Ralpha}\\
    P_{\rm n} = \frac{p_{\rm n,0}}{p_0} P; ~~~P_{\rm c} = \frac{p_{\rm c,0}}{p_0} P.
    \label{eq:Palpha}
\end{gather}

To prevent numerical noise during the early stages of the simulation, we introduce a smoothing function that gradually increases the wave amplitude until it reaches the desired value, $M$.

\section{\label{App:B} The heat flux at the centre of mass frame reference}

The relations between the definitions of the variables in the 1F and 2F models are given by \citep[see, e.g., Appendix D in ][]{KhoColDia2014aa},
\begin{subequations}
    \begin{gather}
        \rho = \sum_{\alpha \in \{\rm n,c\}} \rho_\alpha, \label{eq:rho_t} \\
        \hat{p} = \sum_{\alpha \in \{\rm n,c\}} [\hat{p}_\alpha + \rho_\alpha (\vec{w}_\alpha \otimes \vec{w}_\alpha)], \label{eq:p_tens_t} \\
        p = \sum_{\alpha \in \{\rm n,c\}} \left(p_\alpha + \frac{1}{3}\rho_\alpha w^2_\alpha\right), \label{eq:p_t} \\
        \vec{q} = \sum_{\alpha \in \{\rm n,c\}} \left(\vec{q}_\alpha + \hat{p}_\alpha \cdot \vec{w}_\alpha + \frac{1}{\gamma - 1}p_\alpha \vec{w}_\alpha + \frac{1}{2} \rho_\alpha w_\alpha^2 \vec{w}_\alpha \right), \label{eq:q_t}
        \end{gather}
\end{subequations}
where $\rho$, $\hat{p}$ and, $p$ are, respectively, the density, stress tensor and thermal pressure for the centre of mass frame reference. In addition, we also introduce the heat flux, $\vec{q}$. 

To derive the 1F set of equations from the 2F equations, we proceed by summing up the equations for neutrals and charges and then we apply the definitions above. For the mass and momentum equations, adding \eqref{eq:rho_n} with \eqref{eq:rho_c} and \eqref{eq:mom_n} with \eqref{eq:mom_c} bring, respectively, \eqref{eq:rho} and \eqref{eq:mom}. When we repeat the same process for the energy equation starting from \eqref{eq:eq_e_n} and \eqref{eq:eq_e_c}, the total energy is found to be:
\begin{equation}
    \sum_{\{\rm n,c\}} e_{\alpha} = \sum_{\{\rm n,c\}} \left(\frac{3}{2} p_\alpha + \frac{1}{2} \rho_\alpha u_\alpha^2\right) + p_{\rm m} = p + \frac{1}{2}\rho u^2 + p_{\rm m}, \label{eq:sum_eT}
\end{equation}
and trivially, the RHS of \eqref{eq:eq_e} can be obtained. However, the sum of the flux terms in the energy equation leads to,
\begin{subequations}
\begin{gather}
    \sum_{\{\rm n,c\}} e_{\rm  \alpha} \vec{u}_\alpha = e_{\rm }\vec{u} + p_{\rm m} \vec{w}_{\rm c} \nonumber \\+ \sum_{\{\rm n,c\}} \left(\frac{3}{2} p_\alpha \vec{w}_\alpha + \rho_\alpha \vec{u}\cdot\vec{w}_\alpha \vec{w}_\alpha + \frac{1}{2}\rho_\alpha w^2_\alpha \vec{w}_\alpha\right) \label{eq:sum_eT_u}\\ 
    \sum_{\{\rm n,c\}} \hat{p}_\alpha \cdot \vec{u}_\alpha = \hat{p} \cdot \vec{u} + \sum_{\{\rm n,c\}} (\hat{p}_\alpha \cdot \vec{w}_\alpha -  \rho_\alpha \vec{u}\cdot\vec{w}_\alpha \vec{w}_\alpha) \label{eq:sum_p_u}\\
    \hat{p}_{\rm m}\cdot \vec{u}_{\rm c} = \hat{p}_{\rm m}\cdot \vec{u} + \hat{p}_{\rm m}\cdot\vec{w}_{\rm c} \label{eq:sum_pm}
\end{gather}
\end{subequations}

The combination of the flux terms \eqref{eq:sum_eT_u}, \eqref{eq:sum_p_u} and \eqref{eq:sum_pm} bring out the flux terms present in the 1F energy equation \eqref{eq:eq_e} plus a group of terms which depend on $\vec{w}_\alpha$. By comparing these terms with those in the definition of $\vec{q}$ (see \eqref{eq:q_t}), we find them to be the same. More in detail,
\begin{equation}
\frac{\partial e}{\partial t} + \nabla \cdot \left[e \vec{u} + (\hat{p} + \hat{p}_{\rm m}) \cdot \vec{u} + \vec{S}^* + \vec{q}\right] = \rho \vec{g} \cdot \vec{u}.
    \label{eq:eq_e_qw}
\end{equation}

Therefore, despite our 2F model does not include a heating flux, the associated 1F does include it, but with $\vec{q}_\alpha = 0$. This result reveals that the closures $\vec{q} = 0$ and $\vec{q}_{\alpha} = 0$ are not equivalent, and then they should be checked when comparing the results of both models.

\section{Extended 1F model \label{App:Ext1F}}

If the contributions of $\vec{G}$ (Eq. \eqref{eq:G_novisc}) and $\vec{q}_{\rm w}$ (Eq. \eqref{eq:q1F}) are included in the set of 1F equations presented in section \ref{sec:1F}, the set of 1F equations reads as,
\begin{subequations}
\begin{gather}
    \frac{\partial \rho}{\partial t} + \nabla \cdot (\rho \vec{u}) = 0, \label{eq:rho_ext}\\
    \frac{\partial}{\partial t}(\rho \vec{u}) + \nabla \cdot (\rho \vec{u} \otimes \vec{u} + \hat{p} + \hat{p}_{\rm m}) = \rho \vec{g}, \label{eq:mom_ext}\\
    \frac{\partial e}{\partial t} + \nabla \cdot \left[e \vec{u} + (\hat{p} + \hat{p}_{\rm m}) \cdot \vec{u} + \vec{S}^* + \vec{q}_{\rm w}\right] = \rho \vec{g} \cdot \vec{u},  \label{eq:eq_e_ext} \\
    \frac{\partial \vec{B}}{\partial t} + \nabla \cdot (\vec{u} \otimes \vec{B} - \vec{B} \otimes \vec{u}) = - \nabla \times \vec{E}^*. \label{eq:induc_1F_ext}
\end{gather}
\end{subequations}

Keeping in mind the relations (for further details, see Eqs. \eqref{eq:Estar}, \eqref{eq:q1F} and, \eqref{eq:w_1F})
\begin{gather*}
    \vec{E}^* = - \xi_{\rm n} \vec{w}_{\rm 1F} \times \vec{B}; ~~~\vec{q}_{\rm w} = \frac{\gamma}{\gamma - 1} \xi_{\rm n} (1 - \xi_{\rm n})\,\tilde{\mu}\, p\, \vec{w}_{\rm 1F},\\
    \vec{w}_{\rm 1F} = - \frac{\tilde{\eta}_{\rm A}}{\xi_{\rm n}^2}(\xi_{\rm n} \nabla \cdot \hat{p}_{\rm m} + \vec{G}),
\end{gather*}
it follows,
\begin{subequations}
\begin{gather}
    \vec{E}^* = - \frac{\tilde{\eta}_{\rm A}}{\xi_{\rm n}}\vec{B} \times(\xi_{\rm n} \nabla \cdot \hat{p}_{\rm m} + \vec{G}); ~~~\vec{E}^*_{\rm G} = -\frac{\tilde{\eta}_{\rm A}}{\xi_{\rm n}}\vec{B} \times \vec{G}, \label{eq:Estar2}\\
    \vec{q}_{\rm w} =  - \frac{\gamma}{\gamma - 1} (1 - \xi_{\rm n})\,\tilde{\mu}\, p\, \frac{\tilde{\eta}_{\rm A}}{\xi_{\rm n}}(\xi_{\rm n} \nabla \cdot \hat{p}_{\rm m} + \vec{G}). \label{eq:q_1F_2}
\end{gather}
\end{subequations}

Those are a particular case from the equations presented by, e.g., \citet{ZaqKhoRuc2011aa}, when the terms proportional to $w^2$ and $w^3$ are neglected.

Considering small perturbations for a homogeneous atmosphere, we obtain the following set of linear equations
\begin{subequations}
\begin{gather}
    \frac{\partial \rho_1}{\partial t} + \frac{\partial}{\partial z}(\rho_0 u_{\rm z}) = 0, \label{eq:rho_lin_ext} \\
\rho_0 \frac{\partial u_{\rm z}}{\partial t} + \frac{\partial p_1}{\partial z} =  - \rho_1 g - \frac{B_{\rm x,0}}{\mu_0}\frac{\partial B_{\rm x,1}}{\partial z}, \label{eq:mom_lin_ext} \\
\frac{\partial p_1}{\partial t} + u_{\rm z} \frac{dp_0}{dz} - c_s^2\left(\frac{\partial \rho_1}{\partial t} + u_z \frac{d \rho_0}{dz}\right) + (\gamma - 1) \left(\frac{\partial q_{w,z}}{\partial z}\right)_1 = 0, \label{eq:e_lin_ext} \\
\frac{\partial B_{\rm x,1}}{\partial t} = - B_{\rm x,0} \frac{\partial u_{\rm z}}{\partial z} + c_{\rm A}^2 \rho_0 \frac{\partial}{\partial z}\left(\tilde{\eta}_{\rm A,0} \frac{\partial B_{\rm x,1}}{\partial z}\right) +  \left(\frac{\partial E^*_{\rm G, y}}{\partial z}\right)_1,
\label{eq:induc1F_lin_ext}
\end{gather}
\end{subequations}
where,
\begin{subequations}
    \begin{gather}
    (\gamma - 1) \left(\frac{\partial q_{w,z}}{\partial z}\right)_1 =  - \gamma \xi_{\rm c,0}\tilde{\mu}_0 p_0\tilde{\eta}_{\rm A,0} \left(\frac{B_{\rm x,0}}{\mu_0} \frac{\partial^2 B_{\rm x,1}}{\partial z^2} + \xi_{\rm c,0} \tilde{\mu}_0 \frac{\partial^2 p_1}{\partial z^2}\right), \label{eq:qw_lin}\\
    \left(\frac{\partial E^*_{\rm G, y}}{\partial z}\right)_1 = \xi_{\rm c,0} \tilde{\mu}_0 \tilde{\eta}_{\rm A,0} B_{\rm x,0} \frac{\partial^2 p_1}{\partial z^2}.\label{eq:G_lin}    \end{gather}
\end{subequations}

\end{appendix}
\end{document}